\documentclass{article}[11pt]
\usepackage{authblk}
\usepackage{graphicx}%
\usepackage{multirow}%
\usepackage{amsmath,amssymb,amsfonts}%
\usepackage{amsthm}%
\usepackage{mathrsfs}%
\usepackage[title]{appendix}%
\usepackage{xcolor}%
\usepackage{textcomp}%
\usepackage{manyfoot}%
\usepackage{booktabs}%
\usepackage{algorithm}%
\usepackage{algorithmicx}%
\usepackage{algpseudocode}%
\usepackage{listings}%
\usepackage[margin=1in]{geometry}
\usepackage{bbm}

\theoremstyle{thmstyleone}%
\newtheorem{theorem}{Theorem}

\newtheorem{proposition}[theorem]{Proposition}%

\theoremstyle{thmstyletwo}%

\theoremstyle{thmstylethree}%
\newtheorem{definition}{Definition}%

\begin{document}

\title{Investigation of a two-patch within-host model of hepatitis B viral infection}

%
%
%
%

\author[1,2]{Keoni Castellano}
\author[1,2]{Omar Saucedo}
\author[1,2, *]{Stanca M. Ciupe}
\affil[1]{Department of Mathematics, Virginia Tech, Blacksburg, VA, 24060}
\affil[2]{Virginia Tech Center for the Mathematics of Biosystems, Virginia Tech, Blacksburg, VA, 24060}
\affil[*]{Corresponding author: stanca@vt.edu}

\maketitle
\abstract{Chronic infection with hepatitis B virus (HBV) can lead to formation of abnormal nodular structures within the liver.  To address how changes in liver anatomy affect overall virus-host dynamics, we developed within-host ordinary differential equation models of two-patch hepatitis B infection, one that assumes irreversible and one that assumes reversible movement between nodular structures. We investigated the models analytically and numerically, and determined the contribution of patch susceptibility, immune responses, and virus movement on within-patch and whole-liver virus dynamics. We explored the structural and practical identifiability of the models by implementing a differential algebra approach and the Monte Carlo approach for a specific HBV data set. We determined conditions for viral clearance, viral localization, and systemic viral infection. Our study suggests that cell susceptibility to infection within modular structures, the movement rate between patches, and the immune-mediated infected cell killing have the most influence on HBV dynamics. Our results can help inform intervention strategies. }
\\
\\ 
\emph{Keywords:} Hepatitis B virus, Asymptotic dynamics, Identifiability analysis, Two-patch model, Ordinary differential equations

\section{Introduction}
Chronic hepatitis B, caused by the hepatitis B virus (HBV), remains a major health problem, affecting more than 257 million carriers globally \cite{schweitzer2015estimations, trepo2014hepatitis}. HBV is transmitted vertically from mother to child, through sexual contact, and after exposure with infected blood or body fluids \cite{guidotti2006immunobiology}. The risk of progression from acute to chronic disease is age dependent, with adult immunocompetent patients clearing the virus and infants and children developing chronic infection, followed by liver disease such as cirrhosis and hepatocellular carcinoma \cite{lee1997hepatitis, locarnini2015strategies}. Chronic hepatitis B is rarely cured with available therapeutics \cite{dusheiko2013treatment, pawlotsky2023new}, hence the need for interdisciplinary studies where mathematical modeling and empirical data can merge to better describe the disease evolution following hepatitis B viral infection.

Current mathematical models describing HBV dynamics from acute to chronic infection are based on classical within-host models \cite{nowak1996viral} (with variations to account for hepatitis B virus characteristics), and are mainly validated with HBV DNA titers in the serum of infected patients and chimpanzees \cite{ciupe2007role,ciupe2014antibody, ciupe2007modeling, murray2005dynamics, dahari2009modeling, neumann2010novel, goyal2017role, tsiang1999biphasic, ciupe2024mathematical, carracedo2017understanding, ciupe2021early, forde2016optimal, afrin2025bistability}. Recently, multiscale mathematical models have been developed, with the aim of integrating intracellular aspects of hepatitis B replication into the cellular dynamics \cite{reinharz2021understanding, ciupe2024mathematical, hailegiorgis2023modeling, shekhtman2024modelling, kadelka2021understanding}. They are validated using in-vitro and humanized chimeric or transgenic mice data, and have been instrumental in addressing the role of immune system (both cytolytic killing and non-cytolytic cure) on viral resolution, and the role of viral markers (such as the s- and e-antigens, and the covalently closed circular DNA) on viral persistence \cite{ciupe2011dynamics, shekhtman2024modelling, kadelka2021understanding, ciupe2024mathematical, ciupe2012mathematical}.

\begin{figure}[h!]
\includegraphics[width=.6\linewidth]{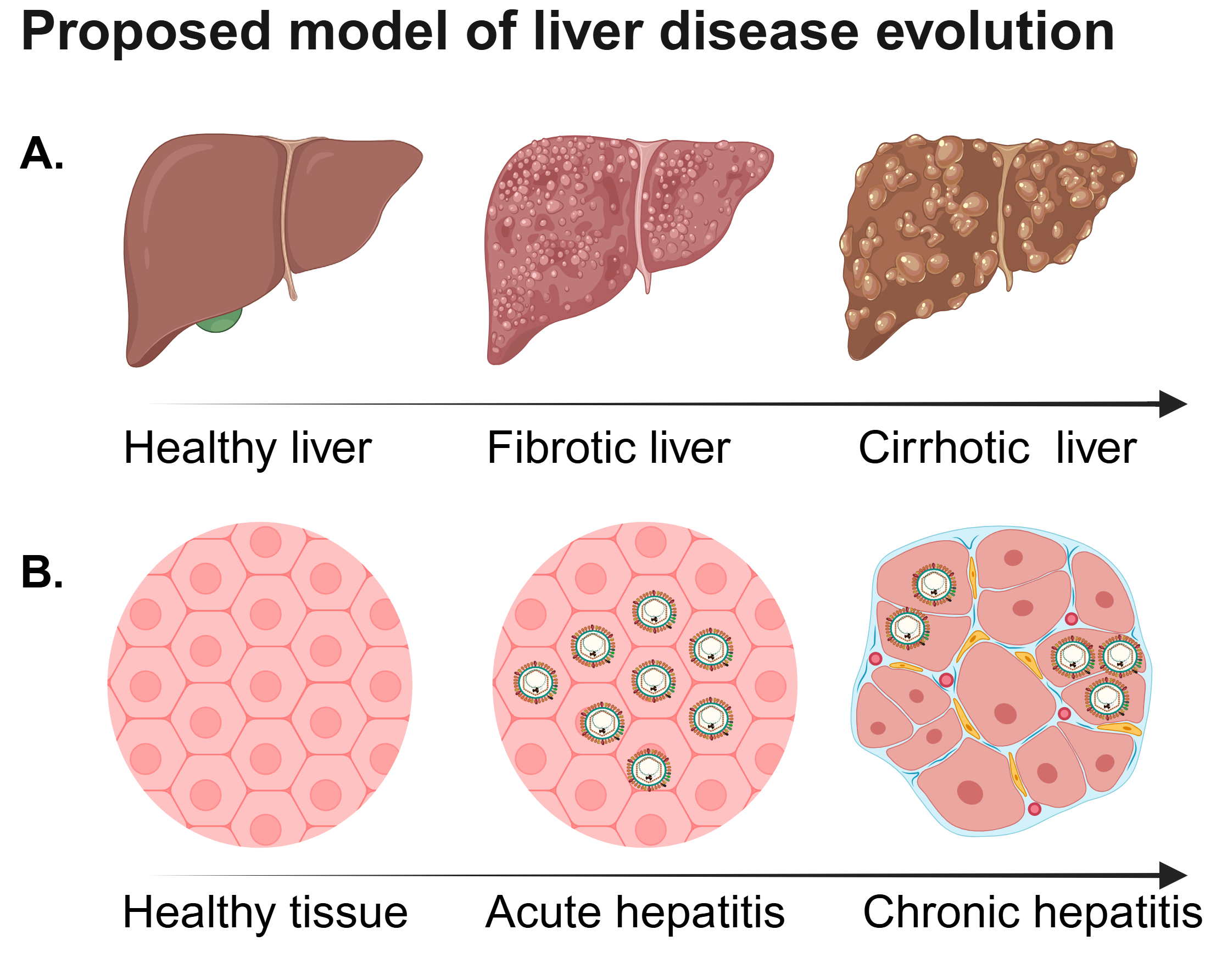}
 \caption{Proposed model of liver disease evolution following hepatitis B viral infection: (\textbf{A.}) Liver transition from healthy to fibrosis to cirrhosis; (\textbf{B.}) Hepatitis B virus localization during acute and chronic stages of hepatitis B viral infection. Figure created in  https://BioRender.com.}
\label{fig:cartoon}
\end{figure}

One aspect of chronic hepatitis B that is less studied, using mathematical models, concerns the effect that hepatitis B virus has on the reshaping of the liver as the disease progresses from healthy, to fibrotic, to cirrhotic (Fig. \ref{fig:cartoon}\textbf{A.}).
The progression to fibrotic and cirrhotic disease is driven by repeated cycles of inflammatory responses, liver cells (hepatocytes) death, hepatocytes regeneration, and accumulation of fibrous tissue \cite{roehlen2020liver, yuan2019hbv}. All these processes lead to the activation of collagen-producing cells, resulting in excessive accumulation of extracellular matrix, the formation of abnormal nodular structures, overall liver dysfunction, and often the need for liver transplantation \cite{bataller2005liver}.

The death of hepatocytes, due to chronic hepatitis B infection, has been shown to trigger compensatory proliferation of the remaining hepatocytes \cite{michalopoulos2021liver}. Fibrosis and cirrhosis affect viral replication, with increased liver stiffness that promotes the progression of hepatitis B disease \cite{cast2019liver}. In this paper, we investigated whether the HBV persistence is influenced by differences in hepatocyte turnover (replication and death) within the nodular strictures formed during fibrosis and cirrhosis. To address this, we developed two-patch mathematical models of heterogeneous hepatitis B virus replication (Fig. \ref{fig:cartoon}\textbf{B.}, right panel) and used them to determine how patch-specific hepatocyte susceptibility, immune response-induced hepatocyte death, and HBV travel rates between patches influence viral persistence. The results can help elucidate how the host-virus interactions in a heterogeneous architecture influence the transition between viral resolution and viral persistence.

\section{Materials and Methods}
\subsection{Mathematical model}
\textbf{Single-patch model of hepatitis B virus infection.} The mathematical model of homogeneous hepatitis B virus infection considers the interaction between uninfected liver cells $T$, infected liver cells $I$, and hepatitis B virus $V$. It assumes that uninfected liver cells are produced at rate $s$, die at natural death rate $d$, and are infected at rate $\beta$. Infected liver cells die at immune-response mediated rate $\delta$ ($\geq d$) and produce new virus at rate $p$. Virus is cleared at rate $c$. The mathematical model is given by:
\begin{equation}\label{model_basic}\begin{split}
    \frac{dT}{dt}&=s- d T -\beta T V,\\
    \frac{dI}{dt}&=\beta T V -\delta I,\\
    \frac{dV_1}{dt}&=p I -c V,\\
\end{split}
\end{equation}
with initial conditions $T_1(0)=s/d$,  $I(0)=I_0$, $V(0)=V_0$. This model was first used (in the context of hepatitis B virus infection) by Nowak et al. \cite{nowak1996viral} and utilized further in \cite{ciupe2007modeling, murray2005dynamics, dahari2009modeling, tsiang1999biphasic, neumann2010novel, ciupe2024mathematical,wang2010global,yousfi2011modeling, meena2023novel, hews2010rich, lewin2001analysis}. Its corresponding basic reproduction number (representing the average number of new viruses or infected cells generated by a single virus in a naive host) is given by
$$R_0=\frac{\beta p}{c\delta}\frac{s}{d}.$$

It is easy to show that virus is cleared when $R_0<1$ and persists when $R_0>1$ \cite{ciupe2017host}. In this study, we expanded model Eq. \ref{model_basic} by taking into account the anatomy of the liver during HBV-induced liver disease. It has been shown that, following hepatitis B infection, liver is restructured into abnormal nodular structures separated by collagen (Fig. \ref{fig:cartoon}\textbf{A.} and \textbf{B.}, right panels). To address differences in virus-host interactions in nodular structures, we developed a two-patch mathematical model and used it to investigate how heterogeneity in infection changes the conditions for viral clearance and viral persistence. Special attention was given to parameters describing liver injury, $\delta$, and liver regeneration, $s$. This two-patch model is a first step in including heterogeneity, with multi-patch models being the obvious next step.\\

\noindent\textbf{Two-patch model of hepatitis B virus infection.} We expanded model Eq. \ref{model_basic} to account for virus dynamics in two liver nodular structures, obtaining a two-patch model of HBV infection. In each nodular structure, we modeled the interaction between uninfected liver cells $T_j$, infected liver cells $I_j$, and hepatitis B virus $V_j$, with $j=\{1,2\}$. As before, we assumed that uninfected liver cells are produced at rates $s_j$ (nodular structure-dependent), die at natural death rate $d$, and get infected at rate $\beta$ (nodular structure-independent). Infected liver cells are killed at rate $\delta$ and produce new virus at rate $p$ (nodular structure-independent). Virus is cleared at rate $c$ (nodular structure-independent). Lastly, we assumed that HBV moves between the nodular structures at rates $\phi_{ij}$ (nodular structure-dependent), for $i,j=\{1,2\}$ and $i\neq j$. The mathematical model becomes:
\begin{equation}\label{model_general}\begin{split}
    \frac{dT_1}{dt}&=s_1- d T_1 -\beta T_1 V_1,\\
    \frac{dI_1}{dt}&=\beta T_1 V_1 -\delta I_1,\\
    \frac{dV_1}{dt}&=p I_1 -c V_1+\phi_{21}V_2-\phi_{12} V_1,\\
    \frac{dT_2}{dt}&=s_2- d T_2 -\beta T_2 V_2,\\
    \frac{dI_2}{dt}&=\beta T_2 V_2-\delta I_2,\\
    \frac{dV_2}{dt}&=p I_2 -c V_2-\phi_{21}V_2+\phi_{12} V_1.\\
\end{split}
\end{equation}

We considered two special cases: (i) the one-directional two-patch model, which assumes that the HBV is seeded in nodular structure 1, moves into nodular structure 2, and never returns into nodular structure 1, given by:

\begin{equation}\label{model_oneD}\begin{split}
    \frac{dT_1}{dt}&=s_1- d T_1 -\beta T_1 V_1,\\
    \frac{dI_1}{dt}&=\beta T_1 V_1 -\delta I_1,\\
    \frac{dV_1}{dt}&=p I_1 -c V_1-\phi V_1,\\
    \frac{dT_2}{dt}&=s_2- d T_2 -\beta T_2 V_2,\\
    \frac{dI_2}{dt}&=\beta T_2 V_2-\delta I_2,\\
    \frac{dV_2}{dt}&=p I_2 -c V_2+\phi V_1,\\
\end{split}
\end{equation}
and (ii) the two-directional two-patch model, which assumes that the HBV is seeded in nodular structure 1, and moves at the same rates between the two patches, given by:
\begin{equation}\begin{split}\label{model_equal}
    \frac{dT_1}{dt}&=s_1- d T_1 -\beta T_1 V_1,\\
    \frac{dI_1}{dt}&=\beta T_1 V_1 -\delta I_1,\\
    \frac{dV_1}{dt}&=p I_1 -c V_1-\phi V_1 + \phi V_2,\\
    \frac{dT_2}{dt}&=s_2- d T_2 -\beta T_2 V_2,\\
    \frac{dI_2}{dt}&=\beta T_2 V_2-\delta I_2,\\
    \frac{dV_2}{dt}&=p I_2 -c V_2+\phi V_1 - \phi V_2.\\
\end{split}
\end{equation}
Both models Eq. \ref{model_oneD} and Eq. \ref{model_equal} have initial conditions $T_1(0)=s_1/d$, $T_2(0)=s_2/d$, $V_1(0)=V_0$, $I_1(0)=I_0$,  $I_2(0)=V_2(0)=0$. This means that the HBV infection starts in nodular structure 1.
\subsection{Structural Identifiability Analysis}
Before validating models Eq. \ref{model_oneD} and Eq. \ref{model_equal} with data, we need to determine if their parameters can be uniquely revealed given unlimited noise-free data. In other words, we determined if the model is \emph{globally structurally identifiable} (for a review regarding structural identifiability see \cite{tuncer2018structural, liyanage2024identifiability, tuncer2025structural, liyanage2025structural}).

Consider a general compartment model:
\begin{equation}\label{dynamicalmodel}
\begin{aligned}
      \dot{\mathbf{x}}(t) &= f\big(\mathbf{}{x}(t),\mathbf{q}\big), \\
      \mathbf{x}(0) &= \mathbf{x}_0, \\
      \mathbf{y}(t) &= \mathbf{g}\big(\mathbf{x}(t),\mathbf{p}\big).
\end{aligned}
\end{equation}
Here,
$$x(t) = \{T_1(t),I_1(t), V_1(t) , T_2(t), I_2(t), V_2(t)\}\in\mathbb{R}^6,$$
are the state variables at time $t$, that solve the ordinary differential equations governed by the rules: $$f\big(\mathbf{}{x}(t),\mathbf{q}\big)\in\mathbb{R}^6,$$ for models Eq. \ref{model_oneD} and Eq. \ref{model_equal}. The parameter vector is:
$$\mathbf{q}=\{s_1, s_2, \beta, d, \delta, c, p, \phi\}\in\mathbb{R}^8,$$
and the initial condition vector is:
$$\mathbf{x}_0\in\mathbb{R}^6.$$
We assumed that the empirical observation represented by $\mathbf{y}(t)\in\mathbb{R}^{12}$ can be explained by models Eq. \ref{model_oneD} and Eq. \ref{model_equal} through function: $$\mathbf{g}\big(\mathbf{x}(t),\mathbf{p}\big)\in\mathbb{R}^{12}.$$

\begin{definition} System Eq. \ref{dynamicalmodel} is said to be globally structurally identifiable for parameter vector $\mathbf{q} $ if, for every other parameter vector $\hat{\mathbf{q}},$  $$y(t,\mathbf{q})=y(t,\mathbf{\hat{q}}) \text{ implies } \mathbf{q}=\mathbf{\hat{q}};$$

\noindent it is locally structurally identifiable for parameter vector $\mathbf{q}$ if, for every other parameter vector $\mathbf{\hat{q}}$, $$y(t,\mathbf{q})=y(t,\mathbf{\hat{q}}), \text{ and } \mathbf{\hat{q}}\in B(\mathbf{q})\text{ implies }  \mathbf{q}=\mathbf{\hat{q}}, \text{ where } B(\mathbf{q}) \text{ is a ball centered at } \mathbf{q};$$
\noindent and it is unidentifiable when at least one of its parameters fails the local identifiability test.
\end{definition}

The methodology for determining the structural identifiability of systems of ordinary differential equations can range from the Taylor series approach \cite{pohjanpalo1978system}, the differential algebra approach \cite{bellu2007daisy,ljung1994global}, the generating series approach \cite{walter1997identification}, the implicit functions approach \cite{xia2003identifiability}, and several others. Additionally, there are several tools (platforms) that can assist in establishing whether a system of ordinary differential equations is identifiable, such as  the  \textit{COMBOS} \cite{meshkat2014finding}, the Differential Algebra for Identifiability of SYstems (\textit{DAISY}, \cite{bellu2007daisy}),  the Exact Arithmetic Rank (\textit{EAR}, \cite{anguelova2012minimal}), the \textit{GenSSI2} \cite{ligon2018genssi}, the Structural Identifiability Analyser (\textit{SIAN}, \cite{hong2019sian}), the \textit{STRIKE-GOLDD} \cite{villaverde2016structural},  and the \textit{StructuralIdentifiability.jl} \cite{dong2023differential}. In this study, we used the differential algebra approach and the DAISY platform \cite{bellu2007daisy} to determine the structural identifiability of systems Eq. \ref{model_oneD} and Eq. \ref{model_equal}. The goal of the structural identifiability is to determine which parameters we can confidently estimated from unlimited observations.

\subsection{Data fitting}
We used previously published longitudinal serum HBV DNA titers from one chimeric mouse with humanized liver \cite{zhang2023replication}. Briefly, a urokinase-type plasminogen activator and severe combined immunodeficient (uPA/SCID) mouse was transplanted human hepatocytes   \cite{tateno2015generation}. Approximately 120–150 days after transplantation (when the growth and proliferation of liver cells is complete), the mouse was infected with hepatitis B virus. Serum HBV DNA was collected during the expansion, peak and persistent stage of HBV infection at days $t_{data}$=\{14, 22, 33, 54,	82,	99,	120, 141, 162, 183,	197, 212\} post inoculation. HBV DNA dynamics reached steady levels, similar to an acute HBV infection that becomes persistent in humans \cite{zhang2023replication}.

\subsubsection{Data fitting procedure for the two-patch  models Eq. \ref{model_oneD} and Eq. \ref{model_equal}}
\indent \textbf{Known parameters.} We assumed that hepatocytes have a life-span of 100 days \cite{wang2017hepatocyte}, resulting in an uninfected hepatocyte death rate of $d=0.01$ /day. Moreover, we assumed that $6.8\times 10^{5}$ hepatocytes/ml are susceptible to HBV infection \cite{ciupe2024mathematical}, resulting in $T(0)=T_1(0)+T_2(0)=6.8\times 10^{5}$ cells/ml and a total recruitment rate $s=d \times T(0)= 6.8\times 10^{3}$ cells/(ml$\times$ day). HBV is cleared at rate $c=4.4$ /day \cite{murray2005dynamics}. Since we fit the model to data from immunosupressed mice, we assumed $\delta=d=0.01$ /day.

We assumed that the HBV inoculum is seeded in patch 1, and set patch 1-specific initial virus and infected cells to $V_1(0)=10^4$ HBV DNA/ml and $I_1(0)=I_0=1$ cells/ml \cite{zhang2023replication}. Moreover, we assumed that there is no virus in patch 2 at the beginning of infection. Therefore, $I_2(0)=0$ cells/ml and $V_2(0)=0$ HBV DNA/ml. We split the number of uninfected hepatocytes between the two patches. $T_1(0)=T_2(0)=3.4\times 10^{5}$ hepatocytes/ml, and assumed three cases for hepatocyte recruitment: (\textbf{case 1}) $s_1=0.1\times s=6.8\times 10^2$ cells/(ml$\times$ day), $s_2=0.9\times s=6.12\times 10^3$ cells/(ml$\times$ day); (\textbf{case 2}) $s_1=s_2=0.5\times s =3.4 \times 10^3$ cells/(ml$\times$ day); and (\textbf{case 3}) $s_1=0.9\times s=6.12\times 10^3$ cells/(ml$\times$ day), $s_2=0.1\times s=6.8\times 10^2$ cells/(ml$\times$ day).

\begin{figure}[h!]
    \centering
\includegraphics[width=.9\linewidth]{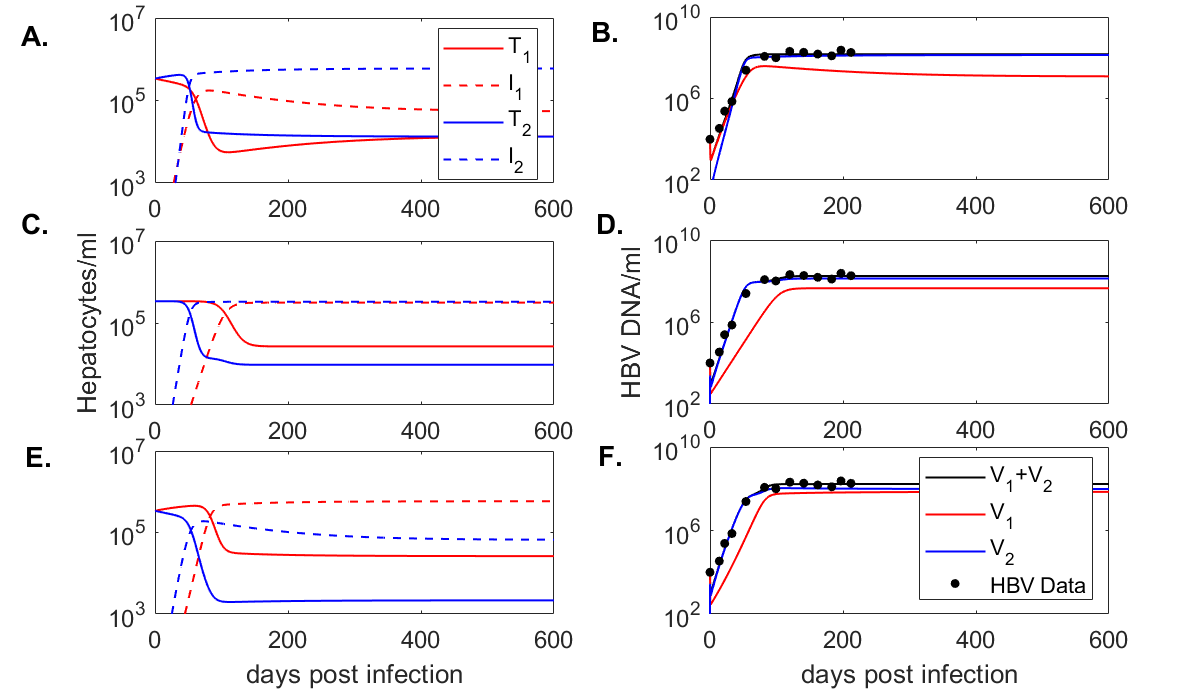}
 \caption{Dynamics of \textbf{(A.)}-\textbf{(E.)} uninfected (solid line) and infected (dashed line) liver cells in patch 1 (red lines) and patch 2 (blue lines) and \textbf{(B.)}-\textbf{(F.)} total virus (black line), virus in patch 1 (red line), virus in patch 2 (blue line) versus mice data (circles) as given by model Eq. \ref{model_oneD} in \textbf{case 1: (A.)}-\textbf{(B.)}; \textbf{case 2: (C.)}-\textbf{(D.)}, \textbf{case 3: (E.)}-\textbf{(F.)}. Model parameters are given in Table \ref{tab:Param_twoPatch_oneDirection}.}
\label{fig:twoPatch_oneDirection}
\end{figure}

\begin{figure}[h!]
    \centering
\includegraphics[width=.9\linewidth]{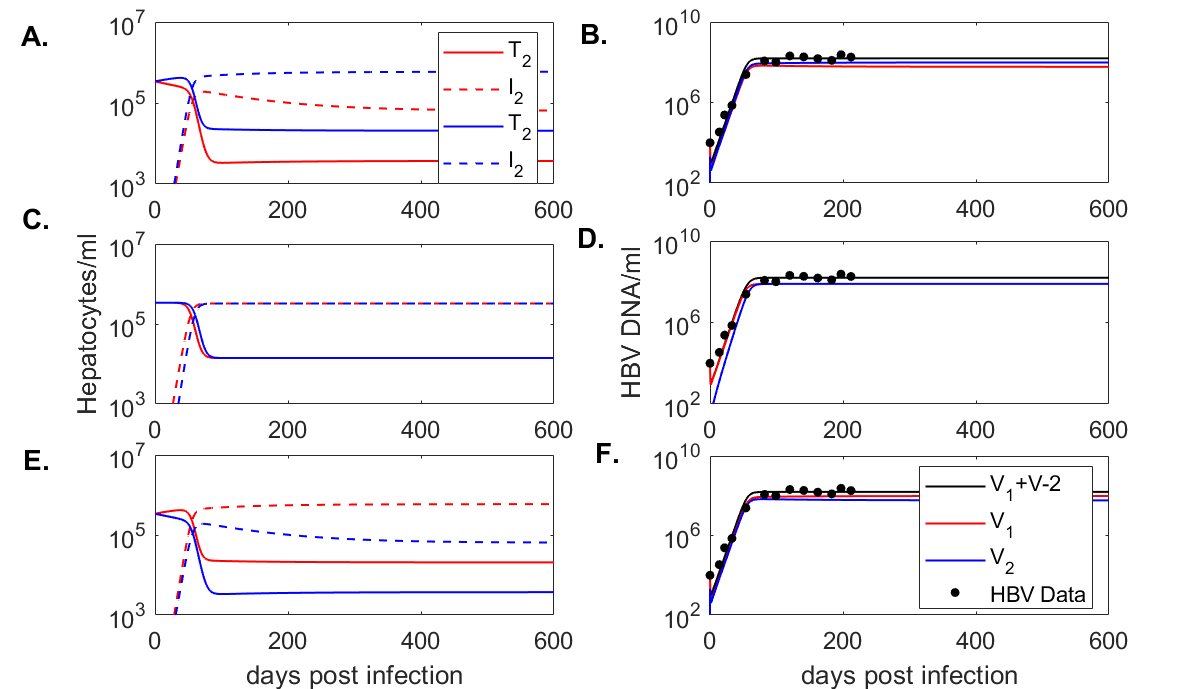}
 \caption{Dynamics of \textbf{(A.)}-\textbf{(E.)} uninfected (solid line) and infected (dashed line) liver cells in patch 1 (red lines) and patch 2 (blue lines) and \textbf{(B.)}-\textbf{(F.)} total virus (black line), virus in patch 1 (red line), virus in patch 2 (blue line) versus mice data (circles) as given by model Eq. \ref{model_equal} in \textbf{case 1: (A.)}-\textbf{(B.)}; \textbf{case 2: (C.)}-\textbf{(D.)}, \textbf{case 3: (E.)}-\textbf{(F.)}. Model parameters are given in Table \ref{tab:Param_twoPatch_twoDirection}.}
\label{fig:twoPatch_twoDirection}
\end{figure}

\textbf{Data fitting algorithm.} The remaining parameters $\pi=\{\beta, p, \phi\}$ are assumed unknown and are estimated by minimizing the functional:
$$J(\pi)=\left(\sum_{t_{data}} \left(\log_{10} V_1(t_{data}, \pi)+\log_{10} V_2(t_{data}, \pi)-\log_{10} V_{data} (t_{data}) \right)^2\right)^{1/2},$$
over the parameter space $\pi$ using   the built-in function \texttt{fminsearchbnd} in MATLAB R2021a. Parameter bounds are $10^{-10}\leq \beta \leq 10^{-7}$  ml/(virion$\times$day), $0.1\leq \phi\leq 5$ /day, and $0\leq p\leq 1500$ virus/(ml$\times$day) for (\textbf{case 1 - case 3}). The initial guesses are $\beta=5\times 10^{-9}$ ml/(virion$\times$day), $\phi=0.5$/day and $p=100$ virus/(ml$\times$day). The resulting values for model Eq. \ref{model_oneD} are given in Table \ref{tab:Param_twoPatch_oneDirection} and the dynamics of model Eq. \ref{model_oneD} over time are given in Fig. \ref{fig:twoPatch_oneDirection}. Similarly, the resulting values for model Eq. \ref{model_equal} are given in Table \ref{tab:Param_twoPatch_twoDirection} and the dynamics of model Eq. \ref{model_equal} over time are given in Fig. \ref{fig:twoPatch_twoDirection}.

\newpage
\clearpage
\begin{table}[h!]
\centering
\scalebox{0.9}{
\begin{tabular}{|l|l|l|l|}
\hline \text {Fixed Parameters all cases} & \text{Description} & \text{Value} & \text{Reference}\\
\hline
$c$&\text{Virus clearance rate}&$4.4$ /day&\cite{murray2005dynamics}\\
$d$&\text{Uninfected hepatocyte death rate}&$0.01$ /day &\cite{wang2017hepatocyte}\\
$\delta$&\text{Killing rate}&$0.01$ /day &\\\hline
 \text {Fixed Parameters \textbf{case 1}} & \text{Description} & \text{Value} & \text{Reference}\\
\hline
$s_1$&\text{Hepatocyte production patch 1}&$6.8 \times 10^{2}$ cells/(ml$\times$ day)& calculated\\
$s_2$&\text{Hepatocyte production patch 2}&6.12 $ \times 10^{3}$ cells/(ml$\times$ day)& calculated\\
\hline
\text {Estimated Parameters \textbf{case 1}} & \text{Description} & \text{Value} & \text{RSS}\\\hline
$\beta$ & \text{Infectivity rate} & $3.3\times 10^{-9}$  ml/(virion$\times$ day) &0.85\\
$p$&\text{Virus production rate} &998 virion/(ml$\times$ day)& -\\
$\phi$&\text{Movement rate}&$0.1$ /day & -\\
\hline
\text {Fixed Parameters \textbf{case 2}} & \text{Description} & \text{Value} & \text{Reference}\\
\hline
$s_1$&\text{Hepatocyte production patch 1}&$3.4 \times 10^{3}$ cells/(ml$\times$ day)& calculated\\
$s_2$&\text{Hepatocyte production patch 2}&$3.4 \times 10^{3}$ cells/(ml$\times$ day)& calculated\\
\hline
\text {Estimated Parameters \textbf{case 2}} & \text{Description} & \text{Value} & \text{RSS}\\\hline
$\beta$ & \text{Infectivity rate} & $2.63\times 10^{-9}$  ml/(virion$\times$ day) &0.69\\
$p$&\text{Virus production rate} & 1203 virion/(ml$\times$ day)& -\\
$\phi$&\text{Movement rate}&$4.1$ /day & -\\
\hline
 \text {Fixed Parameters \textbf{case 3}} & \text{Description} & \text{Value} & \text{Reference}\\
\hline
$s_1$&\text{Hepatocyte production patch 1}&$6.12 \times 10^{3}$ cells/(ml$\times$ day)& calculated\\
$s_2$&\text{Hepatocyte production patch 2}&$6.8 \times 10^{2}$ cells/(ml$\times$ day)& calculated\\ \hline
\text {Estimated Parameters \textbf{case 3}} & \text{Description} & \text{Value} & \text{RSS}\\\hline
$\beta$ & \text{Infectivity rate} & $3.13\times 10^{-9}$  ml/(virion$\times$ day) &0.52\\
$p$&\text{Virus production rate} &1137 virion/(ml$\times$ day)& -\\
$\phi$&\text{Movement rate}&$5$ /day & -\\
\hline \text {Initial conditions} & \text{Description} & \text{Value} & \text{Reference}\\\hline
$T_1(0)$&Uninfected hepatocytes patch 1& $3.4\times 10^{5}$ cells/ml& \cite{ciupe2024mathematical}\\
$T_2(0)$&Uninfected hepatocytes patch 2& $3.4\times 10^{5}$ cells/ml& \cite{ciupe2024mathematical}\\
$I_1(0)$&Infected hepatocytes patch 1& 1 cell/ml& -\\
$I_2(0)$&Infected hepatocytes patch 2& 0 cell/ml& -\\
$V_1(0)$& Inoculum patch 1 & $10^4$ HBV DNA/ml & \cite{zhang2023replication}\\
$V_2(0)$& Inoculum patch 3 & $0$ HBV DNA/ml & -\\
\hline
\end{tabular} }
\caption{Parameter values for model Eq. \ref{model_oneD}.}\label{tab:Param_twoPatch_oneDirection}
\end{table}

\begin{table}[h!]
\centering
\scalebox{0.9}{\begin{tabular}{|l|l|l|l|}
\hline \text {Fixed Parameters all cases} & \text{Description} & \text{Value} & \text{Reference}\\
\hline$c$&\text{Virus clearance rate}&$4.4$ /day&\cite{murray2005dynamics}\\
$d$&\text{Uninfected hepatocyte death rate}&$0.01$ /day &\cite{wang2017hepatocyte}\\
$\delta$&\text{Infected hepatocyte death rate}&$0.01$ /day &\\
\hline \text {Fixed Parameters \textbf{case 1}} & \text{Description} & \text{Value} & \text{Reference}\\
\hline
$s_1$&\text{Hepatocyte production patch 1}&$6.8 \times 10^{2}$ cells/(ml$\times$ day)& calculated\\
$s_2$&\text{Hepatocyte production patch 2}&6.12 $ \times 10^{3}$ cells/(ml$\times$ day)& calculated\\ \hline
\text {Estimated Parameters \textbf{case 1}} & \text{Description} & \text{Value} & \text{RSS}\\\hline
$\beta$ & \text{Infectivity rate} & $2.93\times 10^{-9}$  ml/(virion$\times$ day) &0.81\\
$p$&\text{Virus production rate} &1053 virion/(ml$\times$ day)& -\\
$\phi$&\text{Movement rate}&$5$ /day & -\\
\hline \text {Fixed Parameters \textbf{case 2}} & \text{Description} & \text{Value} & \text{Reference}\\
\hline
$s_1$&\text{Hepatocyte production patch 1}&$3.4 \times 10^{3}$ cells/(ml$\times$ day)& calculated\\
$s_2$&\text{Hepatocyte production patch 2}&$3.4 \times 10^{3}$ cells/(ml$\times$ day)& calculated\\ \hline
\text {Estimated Parameters \textbf{case 2}} & \text{Description} & \text{Value} & \text{RSS}\\\hline
$\beta$ & \text{Infectivity rate} & $2.96\times 10^{-9}$  ml/(virion$\times$ day) &0.76\\
$p$&\text{Virus production rate} &1055 virion/(ml$\times$ day)& -\\
$\phi$&\text{Movement rate}&$0.1$ /day & -\\
\hline \text {Fixed Parameters \textbf{case 3}} & \text{Description} & \text{Value} & \text{Reference}\\
\hline
$s_1$&\text{Hepatocyte production patch 1}&$6.12 \times 10^{3}$ cells/(ml$\times$ day)& calculated\\
$s_2$&\text{Hepatocyte production patch 2}&$6.8 \times 10^{2}$ cells/(ml$\times$ day)& calculated\\ \hline
\text {Estimated Parameters \textbf{case 3}} & \text{Description} & \text{Value} & \text{RSS}\\\hline
$\beta$ & \text{Infectivity rate} & $2.94\times 10^{-9}$  ml/(virion$\times$ day) &0.81\\
$p$&\text{Virus production rate} &1049 virion/(ml$\times$ day)& -\\
$\phi$&\text{Movement rate}& $5$ /day & -\\
\hline \text {Initial conditions} & \text{Description} & \text{Value} & \text{Reference}\\\hline
$T_1(0)$&Uninfected hepatocytes patch 1& $3.4\times 10^{5}$ cells/ml& \cite{ciupe2024mathematical}\\
$T_2(0)$&Uninfected hepatocytes patch 2& $3.4\times 10^{5}$ cells/ml& \cite{ciupe2024mathematical}\\
$I_1(0)$&Infected hepatocytes patch 1& 1 cell/ml& -\\
$I_2(0)$&Infected hepatocytes patch 2& 0 cell/ml& -\\
$V_1(0)$& Inoculum patch 1 & $10^4$ HBV DNA/ml & \cite{zhang2023replication}\\
$V_2(0)$& Inoculum patch 3 & $0$ HBV DNA/ml & -\\
\hline
\end{tabular} }
\caption{Parameter values for model Eq. \ref{model_equal}.}\label{tab:Param_twoPatch_twoDirection}
\end{table}

\newpage
\clearpage

\subsection{Practical Identifiability Analysis}
Practical identifiability is a methodology that considers the noisiness in the data set for a given model, and  performs data fitting under noise considerations.

\begin{definition}
   A model is \textit{practically identifiable} if a unique parameter set can be consistently obtained through the fitting procedure to noisy data.
\end{definition}

There are several approaches and conditions for assessing practical identifiability \cite{wu2008parameter,eisenberg2013identifiability,roosa2019assessing,saucedo2024comparative, tuncer2018structural, ciupe2022identifiability, heitzman2024effect}. In this paper, we used the Monte Carlo (MC) approach to analyze the practical identifiability of the one-directional two-patch model Eq. \ref{model_oneD} and two-directional two-patch model Eq. \ref{model_equal}. The Monte Carlo approach is a sampling technique that uses random numbers and probability distributions to determine the practical identifiability of a model \cite{metropolis1949monte}.
We performed MC simulations by generating $M=1,000$ data sets using the true parameter set  $\pi$ and adding noise to the data in increasing amounts. The MC simulations are outlined in the following steps:
\begin{enumerate}
\item Solve model Eq. \ref{model_oneD} (Eq. \ref{model_equal}, respectively) numerically with the true parameter vector $\pi$ to obtain the output vector $\mathbf{g}\big(\mathbf{x}(t),\pi \big)$ at discrete time points $t_{data}=\{t_i\}_{i=1}^n$.
\item Generate $M=1,000$ data sets with a given measurement error. We assume the error follows a normal distribution with mean $0$ and variance $\sigma^2(t)$; that is, the data are described by: $$\mathbf{y}_{i,j} = g\big(\mathbf{x}(t_i),\pi\big)(1+\epsilon_{i,j}),$$ where $\epsilon_{ij} \sim \mathcal{N}(0,\sigma)$ at the discrete data time points $t_{data}=\{t_i\}_{i=1}^n$ for all $j=\{1,2,...,M\}$ data sets.
\item Estimate the parameter set $\pi_j$, by fitting model Eq. \ref{model_oneD} (model Eq. \ref{model_equal}, respectively) to each of the $M$ simulated data sets. This is achieved by minimizing the difference between model Eq. \ref{model_oneD} (model Eq. \ref{model_equal}, respectively) output and the data generated for the specific scenario:  \[
J(\pi_{j})=\left(\sum_{t_{data}} \left(\log_{10} V_1(t_{data}, \pi_j)+\log_{10} V_2(t_{data}, \pi_j)-\log_{10} V_{data} (t_{data}) \right)^2\right)^{1/2}.
\]
This optimization problem is solved in MATLAB R2021a using the built-in function \texttt{fminsearchbnd}, which is part of the Optimization Toolbox. Since \texttt{fminsearchbnd} is a local solver, the optimized minimum value can be influenced by the starting point. To avoid issues related to the starting value, we use the true parameter values $\pi$ as the initial parameter starting point provided to \texttt{fminsearchbnd}.
\item Calculate the average relative estimation error (ARE) for each parameter in the set $\pi$, as follows:
    \begin{equation}
       \mathrm{ARE} \big(\pi^{(k)}\big)=100 \%  \times \frac{1}{M}  \sum_{j=1}^{M} \frac{\left| \pi^{(k)}- \pi_j^{(k)}\right|}{\left| \pi^{(k)}\right| },
       \label{ARE_eqn}
    \end{equation}
    where $\pi^{(k)}$ is the $k$-th parameter of the  true parameter set $\pi$, and $ \pi_j^{(k)}$ is the $k$-th element of $\pi_j$.
\item Repeat steps 1 through 4, by increasing the measurement error $\sigma = \{0, 1, 5, 10, 20,  30\}\%$.
\end{enumerate}

The objective of this algorithm is to determine if the parameters are sensitive to gradual noise introduced into the given dataset.
We used the AREs as a metric to determine the practical identifiability of model Eq. \ref{model_oneD} (model Eq. \ref{model_equal}, respectively)  by applying the definition below (see  \cite{heitzman2024effect} for details).

\begin{definition}
    Let $\sigma$ be the measurement error introduced to a dataset, and let the ARE be the average measurement error in the parameter $\pi^{(k)}$.
    \begin{itemize}
        \item[1)] If $0\leq$ ARE($\pi^{(k)})\leq\sigma$, we say $\pi^{(k)}$ is strongly practically identifiable.
        \item[2)] If $\sigma<$ ARE($\pi^{(k)})\leq10 \times\sigma$, we say $\pi^{(k)}$ is weakly practically identifiable.
        \item[3)] If $10 \times \sigma<$ARE($\pi^{(k)})$, we say $\pi^{(k)}$ is not practically identifiable.
    \end{itemize}
We state that a model is practically identifiable if $\pi^{(k)}$ is practically identifiable for all values of $k$.
\end{definition}

\section{Results for the one-directional two-patch model Eq. \ref{model_oneD}}

\subsection{Asymptotic analysis results for the one-directional two-patch model Eq. \ref{model_oneD}}
We investigated the long-term behavior of the one-directional two-patch model Eq. \ref{model_oneD}, where HBV is seeded in nodular structure 1, moves into nodular structure 2, and never returns into nodular structure 1. Model Eq. \ref{model_oneD} has three non-negative equilibrium solutions. The infection-free equilibrium, representing viral clearance in both patches, is given by:
\begin{align*}
E_0=(T_1^0,I_1^0,V_1^0,T_2^0,I_2^0,V_2^0) = \left(\frac{s_1}{d},0,0,\frac{s_2}{d},0,0\right),
\end{align*}
and is always biologically realistic.
The one-patch chronic equilibrium, representing virus clearance in nodular structure 1 and persistence in nodular structure 2, is given by:
\begin{align*}
E_1=(T_1^1,I_1^1,V_1^1,T_2^1,I_2^1,V_2^1) = \left(\frac{s_1}{d},0,0,\frac{c \delta }{\beta  p},\frac{c d}{\beta  p}(R_{0}^{\textrm{1D}}-1),\frac{d}{\beta}(R_{0}^{\textrm{1D}}-1)\right),
\end{align*}
where:
\begin{equation}\label{R01D} R_{0}^{\textrm{1D}}=\frac{\beta p s_2}{cd\delta}.\end{equation}
Equilibrium $E_1$ is biologically realistic if and only if $R_{0}^{\textrm{1D}}>1$. Lastly, model Eq. \ref{model_oneD} has two two-patch chronic equilibria,
\begin{equation}\begin{split}\label{E2}
     E_2&=(T_1^2,I_1^2,V_1^2,T_2^2,I_2^2,V_2^2)\\
    &= \left(\frac{\delta(c+\phi)}{p\beta}, \frac{d(c+\phi)}{p\beta}(R_{\textrm{eff}}^{\textrm{1D}}-1),\frac{d}{\beta}(R_{\textrm{eff}}^{\textrm{1D}}-1),\frac{cd\delta (R_0^{\textrm{1D}}+1) + \phi\delta d(R_{\textrm{eff}}^{\textrm{1D}}-1) - \sqrt{D}}{2\beta dp} , \right.\\&\left.\frac{cd\delta (R_0^{\textrm{1D}}-1) - \phi\delta d(R_{\textrm{eff}}^{\textrm{1D}}-1) + \sqrt{D}}{2\beta p \delta},\frac{cd\delta (R_0^{\textrm{1D}}-1) + \phi\delta d(R_{\textrm{eff}}^{\textrm{1D}}-1) + \sqrt{D}}{2\beta c\delta} \right),\\
\end{split}
\end{equation}
and
\begin{equation}\begin{split}
    E_3&=(T_1^3,I_1^3,V_1^3,T_2^3,I_2^3,V_2^3)\\
    &= \left(\frac{\delta(c+\phi)}{p\beta}, \frac{d(c+\phi)}{p\beta}(R_{\textrm{eff}}^{\textrm{1D}}-1),\frac{d}{\beta}(R_{\textrm{eff}}^{\textrm{1D}}-1),\frac{cd\delta (R_0^{\textrm{1D}}+1) + \phi\delta d(R_{\textrm{eff}}^{\textrm{1D}}-1) + \sqrt{D}}{2\beta dp} \right., \\ &\left. \frac{cd\delta (R_0^{\textrm{1D}}-1) - \phi\delta d(R_{\textrm{eff}}^{\textrm{1D}}-1) - \sqrt{D}}{2\beta p \delta},\frac{cd\delta (R_0^{\textrm{1D}}-1) + \phi\delta d(R_{\textrm{eff}}^{\textrm{1D}}-1)- \sqrt{D}}{2\beta c\delta }\right).\\
\end{split}\end{equation}
where
\begin{equation}\label{E2_existence}R_{\textrm{eff}}^{\textrm{1D}}=\frac{\beta p s_1}{d\delta(c+\phi)},\end{equation}
and
\begin{equation*}\begin{split}
    D &=
    (cd\delta)^2(R_0^{\textrm{1D}}-1)^2 + 2c\phi d^2\delta^2(R_0^{\textrm{1D}}+1) (R_{\textrm{eff}}^{\textrm{1D}}-1)+(\phi d \delta)^2 (R_{\textrm{eff}}^{\textrm{1D}}-1)^2.
\end{split}\end{equation*}

Equilibrium $E_2$ is biologically realistic if and only if $\mathcal{R}_{\textrm{eff}}^{\textrm{1D}}>1$. By contrast, $E_3$ is not biologically realistic since $I_2^3$ is always negative. Next, we studied the asymptotic stability of the equilibrium solutions $E_0$, $E_1$ and $E_2$.

\begin{proposition} The infection-free equilibrium $E_0$ is locally asymptotically stable if
\begin{equation*}\label{R0_1D}
    \max\{\mathcal{R}_0^{\textrm{1D}},\mathcal{R}_{\textrm{eff}}^{\textrm{1D}}\}<1,
\end{equation*}
and is unstable otherwise.

\proof
We linearized Eq. \ref{model_oneD} at the disease-free equilibrium $E_0$,
\begin{equation*}
\frac{d}{dt}
    \begin{bmatrix}
        T_1\\I_1\\V_1\\T_2\\I_2\\V_2
    \end{bmatrix}(E_0)=\begin{bmatrix}
        -d & 0 & -\frac{\beta s_1}{d} & 0 & 0 & 0\\
        0 & -\delta & \frac{\beta s_1}{d} & 0 & 0 & 0\\
        0 & p & -c-\phi & 0 & 0 & 0\\0 & 0 & 0 & -d & 0 & -\frac{\beta s_2}{d}\\0 & 0 & 0 & 0 & -\delta & \frac{\beta s_2}{d}\\
        0 & 0 & \phi & 0 & p & -c
    \end{bmatrix}\begin{bmatrix}
        T_1\\I_1\\V_1\\T_2\\I_2\\V_2
    \end{bmatrix}(E_0).
\end{equation*}
The corresponding Jacobian matrix at the disease-free equilibrium $E_0$,
\begin{equation*}
    J(E_0) = \begin{bmatrix}
        -d & 0 & -\frac{\beta s_1}{d} & 0 & 0 & 0\\
        0 & -\delta & \frac{\beta s_1}{d} & 0 & 0 & 0\\
        0 & p & -c-\phi & 0 & 0 & 0\\0 & 0 & 0 & -d & 0 & -\frac{\beta s_2}{d}\\0 & 0 & 0 & 0 & -\delta & \frac{\beta s_2}{d}\\
        0 & 0 & \phi & 0 & p & -c
    \end{bmatrix},
\end{equation*}
can be written as:
\begin{equation*}
    J(E_0) = \begin{bmatrix}
        A_{01} & \mathbf{0}\\
        \Phi & A_{02}
    \end{bmatrix},
\end{equation*}
where $\mathbf{0}$ is the $3\times 3$ zero matrix,
\begin{align*}
    A_{01} &= \begin{bmatrix}
        -d & 0 & -\frac{\beta s_1}{d}\\
        0 & -\delta & \frac{\beta s_1}{d}\\
        0 & p & -c-\phi
    \end{bmatrix},
    A_{02} =\begin{bmatrix}
        -d & 0 & -\frac{\beta s_2}{d}\\
        0 & -\delta & \frac{\beta s_2}{d}\\
        0 & p & -c
    \end{bmatrix},
    \Phi = \begin{bmatrix}
        0 & 0 & 0\\
        0 & 0 & 0\\
        0 & 0 & \phi
    \end{bmatrix}.
\end{align*}
The characteristic equation of model Eq. \ref{model_oneD} at $E_0$ is:
\begin{align*}
    0 = P_0(\lambda)&=\det(J(E_0)-\lambda \mathbbm{1}_6)=\det(A_{01}-\lambda \mathbbm{1}_3)\det(A_{02}-\lambda \mathbbm{1}_3)
    = (\lambda+d)^2P_{01}(\lambda)P_{02}(\lambda).
\end{align*}
where
\begin{align*}
    P_{01}(\lambda) &= \begin{vmatrix}
        -\delta-\lambda & \frac{\beta s_1}{d}\\
        p & -c-\phi-\lambda
    \end{vmatrix}
    = \lambda^2+(c+\delta+\phi)\lambda
    +\delta (c+\phi)(1-R_{\textrm{eff}}^{\textrm{1D}}),\\
    P_{02}(\lambda) &= \begin{vmatrix}
        -\delta-\lambda & \frac{\beta s_2}{d}\\
        p & -c-\lambda
    \end{vmatrix} = \lambda^2+(c+\delta)\lambda + c\delta (1-\mathcal{R}_0^{\textrm{1D}}),
\end{align*}
and $\mathbbm{1}_k$ is the $k\times k$ identity matrix. Two eigenvalues, $\lambda_{1,2}=-d$ are always negative. It is easy to see that the solutions of $P_{01}(\lambda)$ are negative or have negative real parts if $R_{\textrm{eff}}^{\textrm{1D}}<1$ and the solutions of $P_{02}(\lambda)$ are negative or have negative real parts if $\mathcal{R}_0^{\textrm{1D}}<1$. Thus, when $$\max\{\mathcal{R}_0^{\textrm{1D}},R_{\textrm{eff}}^{\textrm{1D}}\}<1,$$ the infection-free equilibrium $E_0$ is locally asymptotically stable. If, by contrast, either $\mathcal{R}_0^{\textrm{1D}}>1$ or $R_{\textrm{eff}}^{\textrm{1D}}>1$, the infection-free equilibrium $E_0$ is unstable. This completes the proof.
\qed
\end{proposition}

\begin{proposition} The one-patch chronic equilibrium $E_1$ exists when $\mathcal{R}_0^{\textrm{1D}}>1$, is locally asymptotically stable if
\begin{equation*}
R_{\textrm{eff}}^{\textrm{1D}} < 1 < \mathcal{R}_0^{\textrm{1D}},
\end{equation*}
and is unstable otherwise.
\proof
We linearize model Eq. \ref{model_oneD} at equilibrium $E_1$,
\begin{equation*}
\frac{d}{dt}
    \begin{bmatrix}
        T_1\\I_1\\V_1\\T_2\\I_2\\V_2
    \end{bmatrix}(E_1)=\begin{bmatrix}
        -d & 0 & -\frac{\beta s_1}{d} & 0 & 0 & 0\\
        0 & -\delta & \frac{\beta s_1}{d} & 0 & 0 & 0\\
        0 & p & -c-\phi & 0 & 0 & 0\\0 & 0 & 0 & -\frac{\beta p s_2}{c\delta} & 0 & -\frac{c\delta}{p}\\0 & 0 & 0 & -d + \frac{\beta p s_2}{c\delta} & -\delta & \frac{c\delta}{p}\\
        0 & 0 & \phi & 0 & p & -c
    \end{bmatrix}\begin{bmatrix}
        T_1\\I_1\\V_1\\T_2\\I_2\\V_2
    \end{bmatrix}(E_1).
\end{equation*}
The corresponding Jacobian matrix at equilibrium $E_1$,
\begin{equation*}
    J(E_1) = \begin{bmatrix}
        -d & 0 & -\frac{\beta s_1}{d} & 0 & 0 & 0\\
        0 & -\delta & \frac{\beta s_1}{d} & 0 & 0 & 0\\
        0 & p & -c-\phi & 0 & 0 & 0\\0 & 0 & 0 & -\frac{\beta p s_2}{c\delta} & 0 & -\frac{c\delta}{p}\\0 & 0 & 0 & -d + \frac{\beta p s_2}{c\delta} & -\delta & \frac{c\delta}{p}\\
        0 & 0 & \phi & 0 & p & -c
    \end{bmatrix},
\end{equation*}
can be written as:
\begin{equation*}
    J(E_1) = \begin{bmatrix}
        A_{11} & \mathbf{0}\\
        \Phi & A_{12}
    \end{bmatrix},
\end{equation*}
where
\begin{align*}
    A_{11} = \begin{bmatrix}
        -d & 0 & -\frac{\beta s_1}{d}\\
        0 & -\delta & \frac{\beta s_1}{d}\\
        0 & p & -c-\phi
    \end{bmatrix},
    A_{12} =\begin{bmatrix}
        -\frac{\beta p s_2}{c\delta}& 0 & -\frac{c\delta}{p}\\
        -d+\frac{\beta p s_2}{c\delta} & -\delta & \frac{c\delta}{p}\\
        0 & p & -c
    \end{bmatrix},
\end{align*}
and $\mathbf{0}$, $\Phi$ are as before (see \textbf{Proposition 1.}). The characteristic equation of model Eq. \ref{model_oneD} at $E_1$ is:
\begin{equation*}
    0 = P_1(\lambda)
    =\det(J(E_1)-\lambda \mathbbm{1}_6)
    = \det(A_{11}-\lambda \mathbbm{1}_3)\det(A_{12}-\lambda \mathbbm{1}_3) = (\lambda +d)P_{11}(\lambda)P_{12}(\lambda),
\end{equation*}
where
\begin{align*}
    P_{11}(\lambda) &= \begin{vmatrix}
        -\delta - \lambda & \frac{\beta s_1}{d}\\
        p & -c-\phi-\lambda
    \end{vmatrix} = \lambda^2+(c+\delta)\lambda + c\delta (1-\mathcal{R}_{\textrm{eff}}^{\textrm{1D}}),\\
    P_{12}(\lambda) &= \begin{vmatrix}
        -\frac{\beta p s_2}{c\delta} - \lambda & 0 & -\frac{c\delta}{p}\\
        -d+\frac{\beta p s_2}{c\delta} & -\delta-\lambda & \frac{c\delta}{p}\\
        0 & p & -c-\lambda
    \end{vmatrix} = \lambda^3 + (c+\delta + \frac{\beta p s_2}{c\delta})\lambda^2+(\frac{\beta p s_2}{\delta}+\frac{\beta p s_2}{c})\lambda +cd\delta(\mathcal{R}_0^{\textrm{1D}}-1).
\end{align*}

One eigenvalue $\lambda=-d$ is always negative. It is easy to show that the eigenvalues of $P_{11}(\lambda)$ are negative or have negative real parts if $R_{\textrm{eff}}^{\textrm{1D}}<1$. For $P_{12}(\lambda)$, we apply the Routh-Hurwitz conditions which state that the solutions are negative or have negative real part if $b_i>0$ for $i=\{1,2,3\}$ and $b_1 b_2>b_3$, where:
\begin{equation*}
    P_{12}(\lambda) = \lambda^3 + b_1\lambda^2 + b_2\lambda + b_3,
\end{equation*}
and
\begin{align*}
    b_1 &= c+\delta + \frac{\beta p s_2}{c\delta},\\
    b_2 &= \frac{\beta p s_2}{\delta} + \frac{\beta p s_2}{c},\\
    b_3 &= cd\delta(\mathcal{R}_0^{\textrm{1D}}-1).
\end{align*}
Observe that $b_1,b_2>0$ always, and $b_3 > 0$ when $\mathcal{R}_0^{\textrm{1D}}>1$. Moreover,
\begin{equation}
    b_1b_2 - b_3 = \left(c+\delta + \frac{\beta p s_2}{c\delta}\right)\left(\frac{\beta p s_2}{\delta} + \frac{\beta p s_2}{c}\right) - cd\delta(\mathcal{R}_0^{\textrm{1D}}-1)\\
     \left(c+ \frac{\beta p s_2}{c\delta}\right)\left(\frac{\beta p s_2}{\delta} + \frac{\beta p s_2}{c}\right)+ \frac{\delta\beta p s_2}{c} + cd\delta > 0.
\end{equation}

Therefore, when $$R_{\textrm{eff}}^{\textrm{1D}} < 1 < \mathcal{R}_0^{\textrm{1D}},$$ the one-patch chronic equilibrium $E_1$ is locally asymptotically stable. Conversely, if $R_{\textrm{eff}}^{\textrm{1D}} > 1$ or $ \mathcal{R}_0^{\textrm{1D}}<1$, the one-patch chronic equilibrium $E_1$ is unstable. This completes the proof.
\qed
\end{proposition}

\begin{proposition}
The biologically realistic two-patch chronic equilibrium $E_2$ exists and is locally asymptotically stable if
$$R_{\textrm{eff}}^{\textrm{1D}}>1.$$
    \proof
    We linearize model Eq. \ref{model_oneD} at equilibrium $E_2$,
{\footnotesize\begin{equation*}
\frac{d}{dt}
    \begin{bmatrix}
        T_1\\I_1\\V_1\\T_2\\I_2\\V_2
    \end{bmatrix}(E_2)=\begin{bmatrix}
        -d-d(R_{\textrm{eff}}^{\textrm{1D}}-1) & 0 & -\frac{\delta (c+\phi)}{p} & 0 & 0 & 0\\
        d(R_{\textrm{eff}}^{\textrm{1D}}-1) & -\delta & \frac{\delta(c+\phi)}{p} & 0 & 0 & 0\\
        0 & p & -c-\phi & 0 & 0 & 0\\0 & 0 & 0 & -\frac{d(R_0^{\textrm{1D}}+1)}{2}- \frac{\phi\delta d(R_{\textrm{eff}}^{\textrm{1D}}-1) + \sqrt{D}}{2c\delta}  & 0 & -\frac{cd\delta (R_0^{\textrm{1D}}+1) + \phi\delta d(R_{\textrm{eff}}^{\textrm{1D}}-1) - \sqrt{D}}{2dp}\\0 & 0 & 0 & \frac{d(R_0^{\textrm{1D}}-1)}{2}+ \frac{\phi\delta d(R_{\textrm{eff}}^{\textrm{1D}}-1) + \sqrt{D}}{2c\delta} & -\delta & \frac{cd\delta (R_0^{\textrm{1D}}+1) + \phi\delta d(R_{\textrm{eff}}^{\textrm{1D}}-1) - \sqrt{D}}{2dp}\\
        0 & 0 & \phi & 0 & p & -c
    \end{bmatrix}\begin{bmatrix}
        T_1\\I_1\\V_1\\T_2\\I_2\\V_2
    \end{bmatrix}(E_2).
\end{equation*}}
The corresponding Jacobian matrix at $E_2$ is,
\begin{equation*}
\begin{split}
    J(E_2) &
    = \begin{bmatrix}
        A_{11} & 0\\
        \Phi & A_{22}
    \end{bmatrix}.
\end{split}
\end{equation*}
where
\begin{align*}
    A_{11} = \begin{bmatrix}
       -d-d(R_{\textrm{eff}}^{\textrm{1D}}-1) & 0 & -\frac{\delta (c+\phi)}{p}\\
        d(R_{\textrm{eff}}^{\textrm{1D}}-1) & -\delta & \frac{\delta(c+\phi)}{p}\\
        0 & p & -c-\phi
    \end{bmatrix},\\
    A_{22} = \begin{bmatrix}
        -\frac{d(R_0^{\textrm{1D}}+1)}{2}- \frac{\phi\delta d(R_{\textrm{eff}}^{\textrm{1D}}-1) + \sqrt{D}}{2c\delta}  & 0 & -\frac{cd\delta (R_0^{\textrm{1D}}+1) + \phi\delta d(R_{\textrm{eff}}^{\textrm{1D}}-1) - \sqrt{D}}{2dp}\\\frac{d(R_0^{\textrm{1D}}-1)}{2}+ \frac{\phi\delta d(R_{\textrm{eff}}^{\textrm{1D}}-1) + \sqrt{D}}{2c\delta} & -\delta & \frac{cd\delta (R_0^{\textrm{1D}}+1) + \phi\delta d(R_{\textrm{eff}}^{\textrm{1D}}-1) - \sqrt{D}}{2dp}\\
        0 & p & -c
    \end{bmatrix},
\end{align*}
and $\mathbf{0}$ and $\Phi$ are as before (see \textbf{Proposition 1}). The characteristic equation of model Eq. \ref{model_oneD} at $E_2$ is:
\begin{align*}
    0=P_2(\lambda)=\det(J(E_2) - \lambda \mathbbm{1}_6) = \det(A_{11} - \lambda \mathbbm{1}_3)\det(A_{22} - \lambda \mathbbm{1}_3)= P_{21}(\lambda) P_{22}(\lambda).
\end{align*}
For both polynomials, we apply the Routh-Hurwitz condition. Let
\begin{equation*}
    P_{21}(\lambda) = \lambda^3 + a_1\lambda^2 + a_2\lambda + a_3,
\end{equation*}
where
\begin{align*}
    a_1 &= c+d+\delta + \phi + d(R_{\textrm{eff}}^{\textrm{1D}}-1),\\
    a_2 &= d(c+\delta +\phi) +d(c+\delta +\phi)(R_{\textrm{eff}}^{\textrm{1D}}-1),\\
    a_3 &= \delta d(c+\phi)(R_{\textrm{eff}}^{\textrm{1D}}-1).
\end{align*}

Polynomial $P_{21}(\lambda)$ has negative roots or roots with negative real parts when $a_i>0$ for all $i$ and $a_1 a_2>a_3$. Note that $a_i>0$ when $R_{\textrm{eff}}^{\textrm{1D}}>1$. Moreover, it is easy to show that $a_1 a_2- a_3>0$ when $R_{\textrm{eff}}^{\textrm{1D}}>1$.

Similarly, let

\begin{equation*}
    P_{22}(\lambda) = \lambda^3 + b_1\lambda^2 + b_2\lambda + b_3,
\end{equation*}
where
\begin{align*}
    b_1 &=c+\delta+ d+\beta (V_2^2) \\&= c+\delta+\frac{d(R_0^{\textrm{1D}}+1)}{2}+\frac{\phi\delta d(R_{\textrm{eff}}^{\textrm{1D}}-1)+\sqrt{D}}{2c\delta},\\
    b_2 &= (T_2^2)\beta p + (V_2^2) \beta c + (V_2^2) \beta\delta + c d + c\delta + d\delta \\ &=\frac{c+\delta}{2c\delta}\left(cd\delta(R_0^{\textrm{1D}}-1)+\phi\delta d(R_{\textrm{eff}}^{\textrm{1D}}-1)+\sqrt{D}\right)+ d\delta + cd+c\delta +\frac{1}{2d}\left(cd\delta(R_0^{\textrm{1D}}+1)+\phi\delta d(R_{\textrm{eff}}^{\textrm{1D}}-1)-\sqrt{D}\right),\\
    b_3 &= (T_2^2) \beta dp + (V_2^2) \beta c\delta + cd\delta\\&=cd\delta (R_0^{\textrm{1D}}+1)+\phi\delta d(R_{\textrm{eff}}^{\textrm{1D}}-1),
\end{align*}
and $V_2^2$ and $T_2^2$ are the  target cell and virus equilibrium values for the nodular structure 2 as given by Eq. \ref{E2}. Polynomial $P_{22}(\lambda)$ has negative roots or roots with negative real parts when $b_i>0$ for all $i$ and $b_1 b_2>b_3$. Note that when $R_{\textrm{eff}}^{\textrm{1D}}>1$, $b_i>0$ for all $i$. Moreover,
$$b_1 b_2-b_3=\beta^2 (c + \delta) V_2^2 + T_2 V_2 \beta^2 p + V_2 \beta (c + \delta)(c+\delta+2d) + T_2\beta p(c + \delta) + (c+\delta)(c+d)(\delta+d),$$
is positive when $R_{\textrm{eff}}^{\textrm{1D}}>1$. Then, by the Routh-Hurwitz condition, $P_{22}(\lambda)$ has negative roots or roots with negative real part when $R_{\textrm{eff}}^{\textrm{1D}}>1$. This guarantees the local stability analysis of the chronic equilibrium $E_2$. Conversely, when $R_{\textrm{eff}}^{\textrm{1D}}<1$, equilibrum $E_2$ does not exist. This concludes the proof.
    \qed
\end{proposition}

\subsection{Structural identifiability results for model Eq. \ref{model_oneD}}

We investigated the structural identifiability of model Eq. \ref{model_oneD} using the differential algebra approach and the DAISY platform \cite{bellu2007daisy} (see the \textbf{Materials and Methods} section for further details) under two assumptions: (i) viral load is measured in both patches and the initial conditions of model Eq. \ref{model_oneD} are either known or unknown; and (ii) susceptible liver cells are measured in both patches and the initial conditions of model Eq. \ref{model_oneD} are either known or unknown.

\begin{proposition}
    When measurements for the $V_1(t)$ and $V_2(t)$ variables are given, system Eq. \ref{model_oneD} is unidentifiable under unknown initial conditions. In particular, parameters $\pi_1=\{\beta, c, d, \delta, \phi \}$ are globally identifiable and parameters $\pi_2=\{s_1, s_2, p\}$ are unidentifiable. If, additionally, initial conditions are known, then all parameters are globally structurally identifiable.
\proof
The input-output equation for model Eq. \ref{model_oneD}, given data for $y_1=V_{1}(t)$ and $y_2=V_{2}(t)$, is:

\begin{align*}
 &f(y_1, y_2, \pi_1, \pi_2)=\frac{d^3 y_1}{dt^3} y_1 - \frac{d^2 y_1}{dt^2} \frac{d y_1}{dt} + \frac{d^2 y_1}{dt^2} y_1^2 \beta + \frac{d^2 y_1}{dt^2} y_1 (c + d + \delta + \phi)  - \left( \frac{d y_1}{dt} \right)^2 (c + \delta + \phi) \\&+ \frac{d y_1}{dt} y_1^2 \beta (c + \delta + \phi)  + \frac{d y_1}{dt} y_1 d (c + \delta + \phi) + y_1^3 \beta \delta (c + \phi) + y_1^2 \left( - \beta p s_1 + c d \delta + d \delta \phi \right)=0,\\
&g(y_1, y_2, \pi_1, \pi_2)=\frac{d^2 y_1}{dt^2} y_2 \phi + \frac{d y_1}{dt} \frac{d y_2}{dt} \phi - \frac{d y_1}{dt} y_2^2 \beta \phi - \frac{d y_1}{dt} y_2 \phi (d + \delta) + \frac{d^3 y_2}{dt^3} y_2 - \frac{d^2 y_2}{dt^2} \frac{d y_2}{dt} + \frac{d^2 y_2}{dt^2} y_2^2 \beta \\
&+ \frac{d^2 y_2}{dt^2} y_2 (c + d + \delta) - \left( \frac{d y_2}{dt} \right)^2 (c + \delta) + \frac{d y_2}{dt} y_1 \delta \phi + \frac{d y_2}{dt} y_2^2 \beta (c + \delta) \\
&+ \frac{d y_2}{dt} y_2 d (c + \delta) - y_1 y_2^2 \beta \delta \phi - y_1 y_2 d \delta \phi + y_2^3 \beta c \delta + y_2^2 \left( - \beta p s_2 + c d \delta \right)=0.
\end{align*}

Parameters $\{\pi_1,\pi_2\}$ are globally structurally identifiable if:
$$f(y_1, y_2, \pi_1, \pi_2)=f(y_1, y_2, \hat{\pi}_1, \hat{\pi}_2) \textrm{ implies } \pi_1=\hat{\pi}_1 \textrm{ and } \pi_2=\hat{\pi}_2.$$
It is easy to see that:
$$\{\beta=\hat{\beta}\} \textrm{ , } \{d=\hat{d}\} \textrm{ , } \{\delta=\hat{\delta}\} \textrm{ , } \{c=\hat{c}\} \textrm{ , } \{\phi=\hat{\phi}\} \textrm{ , } \{p \times s_1=\hat{p}\times\hat{s}_1\} \textrm{ , } \{p \times s_2=\hat{p}\times\hat{s}_2\}.$$
Therefore $\{\beta, c, d, \delta, \phi\}$ are globally structurally identifiable and $\{s_1, s_2, p\}$ are unidentifiable. When initial conditions are known, we have:
$$\frac{dV_{1}}{dt}(0)=p I_1(0) -cV_1(0)-\phi V_1(0)+\phi V_2(0).$$
As long as $I_1(0)\neq 0$, $p$ (and by default $s_1$ and $s_2$) are globally structurally identifiable. Hence, model Eq. \ref{model_oneD} is globally structurally identifiable for known initial conditions. A summary of the results is given in Table \ref{tab:Identifiability_onePatch}.
\qed
\end{proposition}

\begin{proposition}
When measurements for the $T_1(t)$  and $T_2(t)$ variables are given, system Eq. \ref{model_oneD} is unidentifiable when initial conditions are unknown. In particular, parameters $\pi_1=\{ c, d, \delta, \phi, s_{1}, s_{2}\}$ are globally structurally identifiable, and parameters $\pi_2=\{\beta, p\}$ are unidentifiable. If, additionally, initial conditions are known, then all parameters are globally structurally identifiable.

\proof
The input-output equations for model Eq. \ref{model_oneD}, given data for $y_1=T_{1}(t)$ and $y_2=T_{2}(t)$, are:
\begin{align*}
 & f(y_1,y_2, \pi_1, \pi_2)= - \frac{d^3 y_1}{dt^3} y_1^2 y_2^2 + 3 \frac{d^2 y_1}{dt^2} \frac{d y_1}{dt} y_1 y_2^2  - \frac{d^2 y_1}{dt^2} y_1^2 y_2^2 (c + \delta + \phi) - \frac{d^2 y_1}{dt^2} y_1 y_2^2 s_1
 - 2 \left( \frac{d y_1}{dt} \right)^3 y_2^2 \\
&+ \left( \frac{d y_1}{dt} \right)^2 y_1 y_2^2 (c + \delta + \phi)  + 2 \left( \frac{d y_1}{dt} \right)^2 y_2^2 s_1 + \frac{d y_1}{dt} y_1^3 y_2^2 \beta p
 - \frac{d y_1}{dt} y_1^2 y_2^2 \delta (c + \phi) - \frac{d y_1}{dt} y_1 y_2^2 s_1 (c + \delta + \phi) \\
&
 + \frac{d^2 y_2}{dt^2} y_1^3 y_2 \phi - \left( \frac{d y_2}{dt} \right)^2 y_1^3 \phi  + \frac{d y_2}{dt} y_1^3 y_2 \delta \phi + \frac{d y_2}{dt} y_1^3 \phi s_2
 + y_1^4 y_2^2 \beta d p - y_1^3 y_2^2 \left( \beta p s_1 + c d \delta \right)
 - y_1^3 y_2 \delta \phi s_2 \\
&+ y_1^2 y_2^2 \delta s_1 (c + \phi)=0,\\
&g(y_1,y_2, \pi_1, \pi_2) = \frac{d^2 y_1}{dt^2} y_1 y_2^3 \phi - \left( \frac{d y_1}{dt} \right)^2 y_2^3 \phi + \frac{d y_1}{dt} y_1 y_2^3 \delta \phi + \frac{d y_1}{dt} y_2^3 \phi s_1 - \frac{d^3 y_2}{dt^3} y_1^2 y_2^2 + 3 \frac{d^2 y_2}{dt^2} \frac{d y_2}{dt} y_1^2 y_2 \\
&- \frac{d^2 y_2}{dt^2} y_1^2 y_2^2 (c + \delta) - \frac{d^2 y_2}{dt^2} y_1^2 y_2 s_2 - 2 \left( \frac{d y_2}{dt} \right)^3 y_1^2 + \left( \frac{d y_2}{dt} \right)^2 y_1^2 y_2 (c + \delta)
+ 2 \left( \frac{d y_2}{dt} \right)^2 y_1^2 s_2 \\
&+ \frac{d y_2}{dt} y_1^2 y_2^3 \beta p - \frac{d y_2}{dt} y_1^2 y_2^2 c \delta - \frac{d y_2}{dt} y_1^2 y_2 s_2 (c + \delta)+ y_1^2 y_2^4 \beta d p + y_1^2 y_2^3 \left( - \beta p s_2 - c d \delta + d \delta \phi \right) + y_1^2 y_2^2 c \delta s_2 \\
&- y_1 y_2^3 \delta \phi s_1=0.
\end{align*}

Parameters $\{\pi_1,\pi_2\}$ are globally structurally identifiable if:
$$f(y_1, y_2, \pi_1, \pi_2)=f(y_1, y_2, \hat{\pi}_1, \hat{\pi}_2) \textrm{ and } g(y_1, y_2, \pi_1, \pi_2)=g(y_1, y_2, \hat{\pi}_1, \hat{\pi}_2) \textrm{ imply } \pi_1=\hat{\pi}_1 \textrm{ and } \pi_2=\hat{\pi}_2.$$
It is easy to see that:
$$\{s_1=\hat{s}_1\} \textrm{ , } \{s_2=\hat{s}_2\} \textrm{ , } \{\delta=\hat{\delta}\} \textrm{ , } \{d=\hat{d}\} \textrm{ , } \{c=\hat{c}\} \textrm{ , } \{\phi=\hat{\phi}\} \textrm{ , } \{\beta\times p=\hat{\beta}\times \hat{p}\}.$$
Therefore $\{s_1, s_2, d, \delta, c, \phi\}$ are globally structurally identifiable and $\{\beta, p\}$ are unidentifiable. When initial conditions are known, we have:
$$\frac{dT_{1}}{dt}(0)=s_1-d T_1(0)-\beta T_1(0) V_1(0),$$
making $\beta$ (and by default $p$) globally structurally identifiable. Hence, model Eq. \ref{model_oneD} is globally structurally identifiable for known initial conditions. A summary of the results is given in Table \ref{tab:Identifiability_onePatch}.
\qed
 \end{proposition}

\begin{table}[h!]
\centering
\begin{tabular}{|p{1.6in} | p{1.7in}| p{2in}|}
  \hline
  Observe States & Initial Conditions Known & Initial Conditions Unknown\\ \hline

Model with $V_{1}(t)$ and $V_{2}(t)$ data & Globally structurally identifiable \{$\beta$, c, d, $\delta$, p, $\phi$, $s_{1}$, $s_{2}$\} & Globally structurally identifiable \{c,d,$\beta$,$\phi$,$\delta$\}; Unidentifiable \{$s_1$, $s_2$, p\};  Correlations \{$s_1\times p=\hat{s}_1\times \hat{p}\}$, \{$s_2\times p=\hat{s}_2\times \hat{p}$\}
\\ \hline

Model with $T_{1}(t)$ and $T_{2}(t)$ data &Globally structurally identifiable \{$\beta$, c, d, $\delta$, p, $\phi$, $s_{1}$, $s_{2}$\}  & Globally structurally identifiable \{c, d, $\delta$, $\phi$, $s_{1}$, $s_{2}$\}; Unidentifiable \{$\beta$, p\}; Correlations \{$\beta\times p=\hat{\beta}\times \hat{p}$\} \\ \hline
\end{tabular}
\caption{Identifiability analysis for model Eq. \ref{model_oneD}, performed using the DAISY software \cite{bellu2007daisy}. For the simulations with initial conditions, all initial conditions are known.} \label{tab:Identifiability_onePatch}
\end{table}

\subsection{Numerical results for model Eq. \ref{model_oneD}}
As seen above, under known initial conditions and unlimited noise-free measurements of both $V_1(t)$ and $V_2(t)$ (or $T_1(t)$ and $T_2(t)$), all parameters of model Eq. \ref{model_oneD} can be identified from data.

We assumed that the entire hepatocyte population $s=d\times (T_1(0)+T_2(0))$ is susceptible to HBV infection and considered three cases for the susceptible cells ratio within the two patches $s_1:s_2 =\{10:90, 50:50, 90:10\}$ (see Table \ref{tab:Param_twoPatch_oneDirection}). Since we fitted the model to data from immunosupressed mice \cite{zhang2023replication}, we set $\delta=d=0.01$ /day \cite{ciupe2007modeling} and fixed $c=4.4$ /day, as in prior studies \cite{murray2005dynamics}. We fitted the remaining parameters $\{\beta, p, \phi\}$ using the \texttt{fminsearchbnd} function in Matlab2021a (see \textbf{Materials and Methods} section for details). A summary of estimates for each case is given in Table \ref{tab:Param_twoPatch_oneDirection} and the model dynamics are plotted in Fig. \ref{fig:twoPatch_oneDirection}.

Model fitting resulted in similar infectivity rates among the three patch recruitment cases, \textit{i.e.} $\beta=3.3\times 10^{-9}$ ml/(virion $\times$ day), $\beta=2.63\times 10^{-9}$ ml/(virion $\times$ day), and $\beta=3.13\times 10^{-9}$ ml/(virion $\times$ day), for \textbf{cases 1, 2} and \textbf{3}, respectively. Viral production rate increased by $21\%$ and $14\%$ in \textbf{cases 2} and \textbf{3} compared to \textbf{case 1}, \textit{i.e.}  $p=1203$ virion/(ml$\times$ day) and $p=1137$ virion/(ml$\times$ day), versus $p=998$ virion/(ml$\times$ day). The biggest difference is in the estimate of the movement rate $\phi$, which was 41- and 50-times higher in \textbf{cases 2} and \textbf{3}, compared to \textbf{case 1}, $\phi=4.1$ /day and $\phi=5$ /day, versus $\phi=0.1$ /day. To determine if the results can be generalized when we add noise to the data, we performed practical identifiability analysis using Monte Carlo approaches (see \textbf{Materials and Methods} for details). We found that all parameters are strongly practically identifiable for all cases with the exception of parameter $\phi$, which is weakly practically identifiable for \textbf{case 1} and \textbf{case 2}; and parameter $p$, which is weakly practically identifiable for \textbf{case 1} (see Table \ref{tab:Practical_OneDir}).

\begin{table}[h!]
\centering
\begin{tabular}{|c||c|c|c||c|c|c||c|c|c||}
\hline
\multicolumn{1}{|c||}{} & \multicolumn{3}{|c||}{\textbf{case 1}} & \multicolumn{3}{|c||}{\textbf{case 2}} & \multicolumn{3}{|c||}{\textbf{case 3}} \\
\hline ARE & $\beta$ &$\phi$  & $p$ & $\beta$ &$\phi$  & $p$ & $\beta$ &$\phi$  & $p$ \\
\hline 0\% & 0 & 0 & 0 & 0 & 0 & 0 & 0 & 0 & 0 \\
\hline 1\% & 0.3502 & 1.8354 & 4.7200 & 0.0542 & 12.5492 & 0.5759 & 0.0313 & 0.6557 & 0.4824 \\
\hline 5\% & 0.4568 & 24.4316 & 6.3968 & 0.0700 & 12.7504 & 0.9358 & 0.1293 & 1.2374 & 2.0864 \\
\hline 10\% & 0.5899 & 36.8748 & 8.3112 & 0.1391 & 13.2641 & 2.0531 & 0.2481 & 2.2095 & 4.0030 \\
\hline 20\% & 0.7314 & 45.5693 & 10.6981 & 0.3188 & 13.1036 & 4.8778 & 0.4282 & 3.0876 & 6.9326 \\
\hline 30\% & 0.9340 & 57.2279 & 13.8884 & 0.6018 & 13.3719 & 8.8351 & 0.7088 & 3.4158 & 10.8375 \\
\hline Identifiable? & Yes & Weakly & Weakly & Yes & Weakly & Yes & Yes & Yes & Yes    \\
\hline
\end{tabular}
\caption{MC approach for the one-directional two-patch model Eq. \ref{model_oneD}.}  \label{tab:Practical_OneDir}
\end{table}

In all three cases we are in the $R_0^{1D}>1$ and $R_{eff}^{1D}>1$ regime ($R_0^{1D}=45.8$, $R_{eff}^{1D}=4.9$ for \textbf{case 1}; $R_0^{1D}=24.4$, $R_{eff}^{1D}=12.6$ for \textbf{case 2}; and $R_0^{1D}=5.5$, $R_{eff}^{1D}=23.1$ for \textbf{case 3}), which means virus persists in both patches (see \textbf{Proposition 5}).

\begin{figure}[h!]
    \centering
    \includegraphics[scale = 0.35]{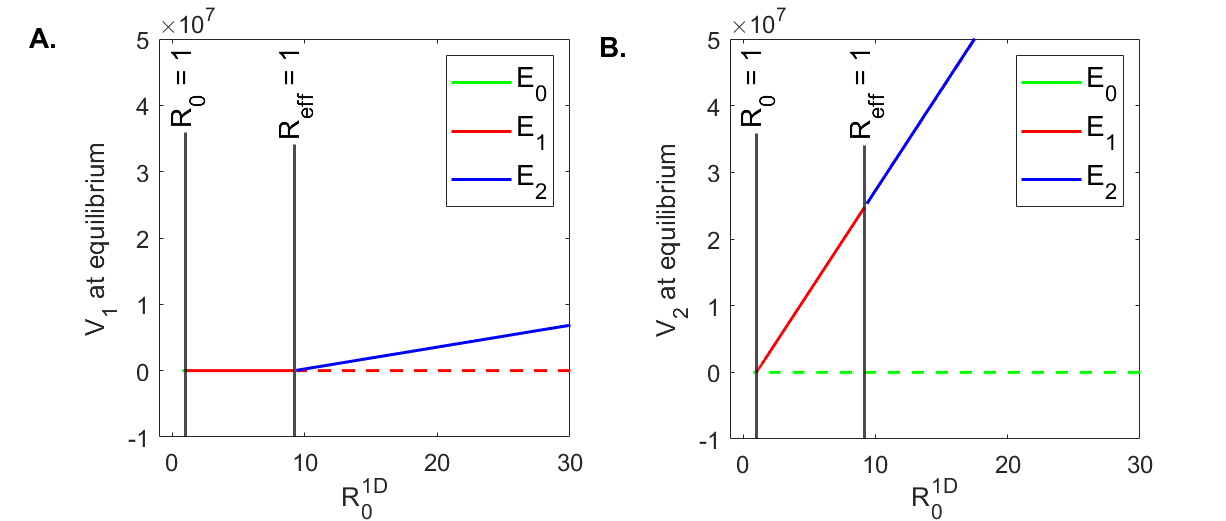} \includegraphics[scale = 0.35]{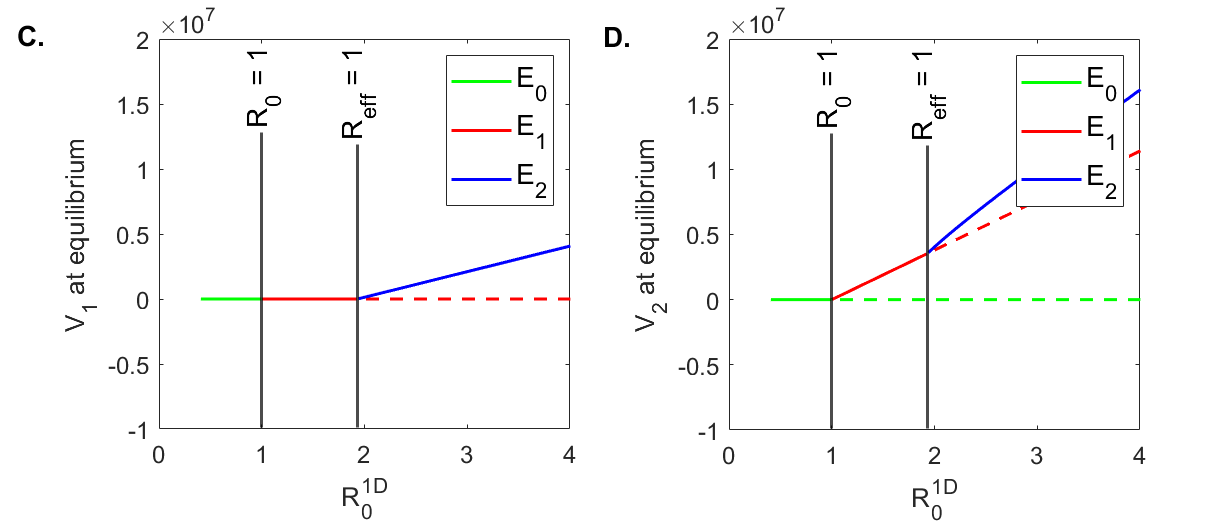} \includegraphics[scale = 0.35]{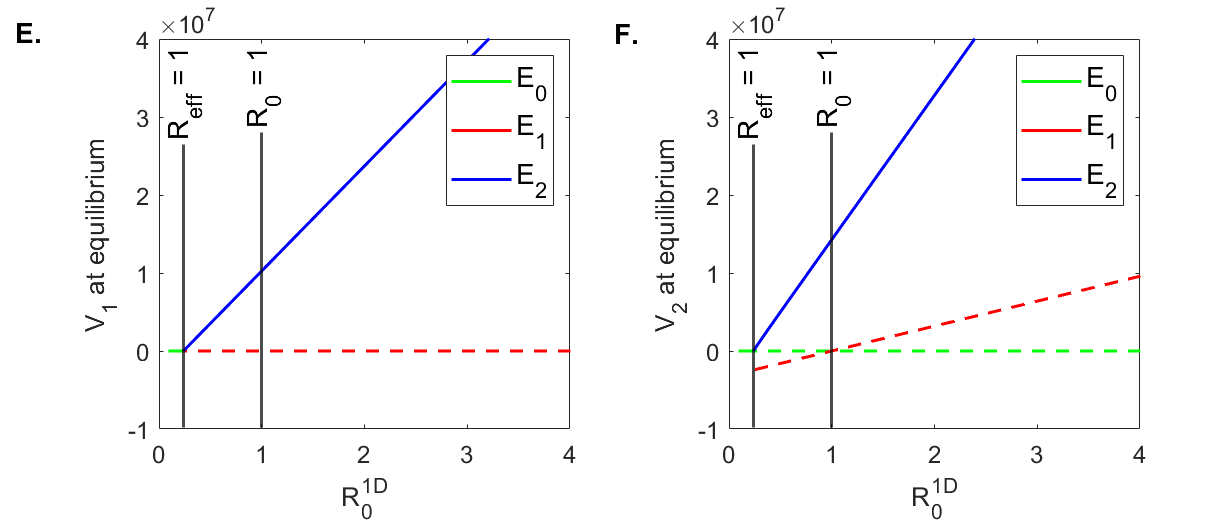}
    \caption{One dimensional bifurcation diagrams for model Eq. \ref{model_oneD}: $V_1$ and $V_2$ at equilibrium versus $R_0^{1D}$. The other parameters are given in Table \ref{tab:Param_twoPatch_oneDirection} for \textbf{A. - B.} \textbf{case 1}, \textbf{C. - D.} \textbf{case 2} and \textbf{E. - F.} \textbf{case 3}.}
    \label{fig:1DBif_oneDir}
\end{figure}

\begin{figure}[h!]
    \centering
    \includegraphics[width=0.32\linewidth]{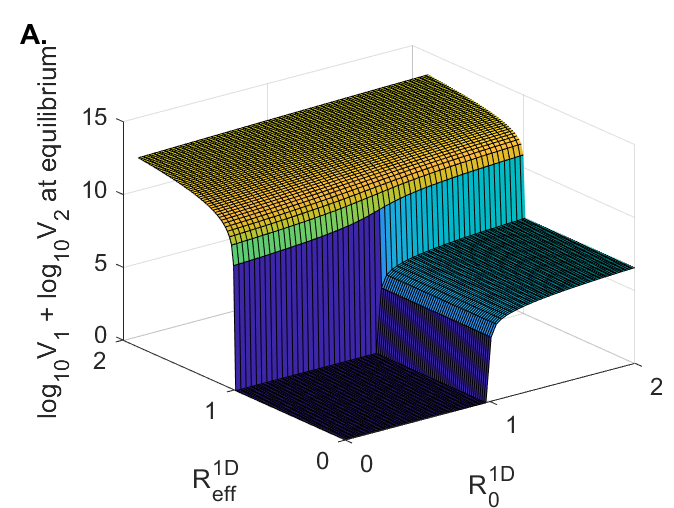} \includegraphics[width=0.32\linewidth]{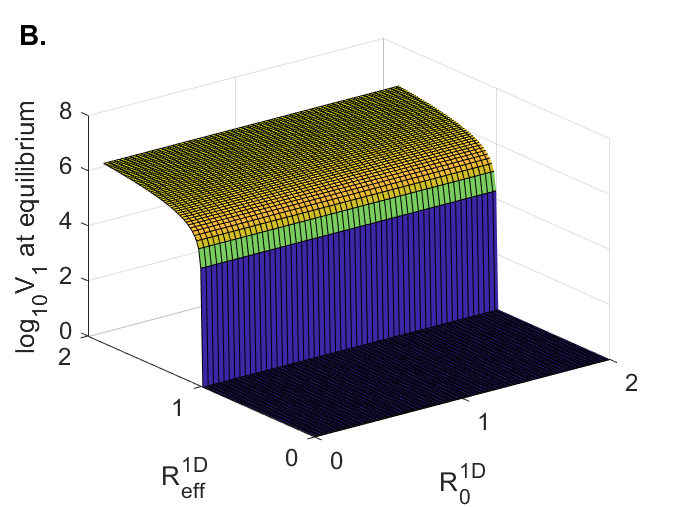}
    \includegraphics[width=0.32\linewidth]{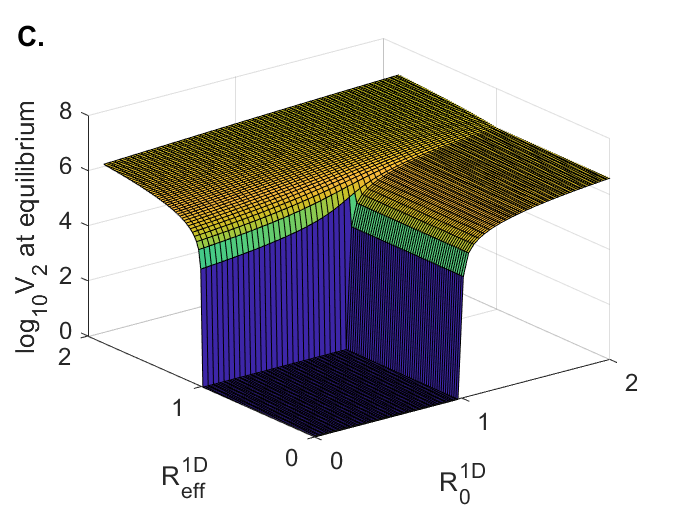}

    \caption{Two dimensional bifurcation diagrams for model Eq. \ref{model_oneD}: \textbf{A.} $\log_{10} V_1+\log_{10} V_2$ at equilibrium; \textbf{B.} $\log_{10} V_1$ at equilibrium; \textbf{C.} $\log_{10} V_2$ at equilibrium versus $R_{eff}^{1D}$ and $R_0^{1D}$. The other parameters are given in Table \ref{tab:Param_twoPatch_oneDirection}, case 2.}
    \label{fig:2D_oneDir}
\end{figure}

To determine conditions where virus is cleared in one or both patches, we created one-dimensional bifurcation diagrams. We used the formulas for $V_1$ and $V_2$ in the equilibrium solutions $E_0$, $E_1$, and $E_2$ (see \textbf{Section 3.1} for detail) and varied values of $\delta\in(0.01, 0.6)$. All results are presented as $V_1$ and $V_2$ at equilibria versus $R_0^{1D}$ (which is inversely proportional to $\delta$). We found that when the number of susceptible cells in patch 1 is low, $s_1:s_2=10:90$ (as in \textbf{case 1}), virus is cleared in patch 1 for a large parameter range $1<R_0^{1D}<9$ (Fig. \ref{fig:1DBif_oneDir}\textbf{A.}) and persists in patch 2 for all $1<R_0^{1D}$ (Fig. \ref{fig:1DBif_oneDir}\textbf{B.}). For equal number of susceptible cells in the two patches, $s_1:s_2=50:50$ (as in \textbf{case 2}), virus is cleared in patch 1 for a smaller parameter range $1<R_0^{1D}<1.7$ (Fig. \ref{fig:1DBif_oneDir}\textbf{C.}) and persists in patch 2 for all $1<R_0^{1D}$ (Fig. \ref{fig:1DBif_oneDir}\textbf{D.}). Lastly, when the number of susceptible cells in patch 1 is high, $s_1:s_2=90:10$ (as in \textbf{case 3}), virus persists in both patches (Fig. \ref{fig:1DBif_oneDir}\textbf{E.} and \textbf{F.}). As shown analytically, virus is cleared from both patches when $R_0^{1D}<1$ and $R_{eff}^{1D}<1$ (Fig. \ref{fig:1DBif_oneDir}, green solid lines). Interestingly, the  long-term results for \textbf{case 1} and \textbf{case 2} have $R_0^{1D}<R_{eff}^{1D}$ (Fig. \ref{fig:1DBif_oneDir}\textbf{A.} - \textbf{D.}) and long-term results for \textbf{cases 3} have $R_0^{1D}>R_{eff}^{1D}$ (Fig. \ref{fig:1DBif_oneDir}\textbf{E.} - \textbf{F.}). To unify our results, we created two-dimensional bifurcation diagrams (using the formulas for the equilibrium values for $V_1$ and $V_2$ of the equilibrium solutions $E_0$, $E_1$, and $E_2$, as before) and varied values of $s_1\in (0, 500)$ and $s_2\in (0, 500)$. We computed $V_1$, $V_2$ and $V=V_1+V_2$ at equilibria versus $R_0^{1D}$ and $R_{eff}^{1D}$ values. As shown analytically, we find that the clearance of virus in patch 1 occurs when $R_{eff}^{1D}<1$, regardless of $R_0^{1D}$ value (Fig. \ref{fig:2D_oneDir}\textbf{B.}), while clearance in patch 2 (and overall) requires both $R_{eff}^{1D}<1$ and $R_0^{1D}<1$ (Fig. \ref{fig:2D_oneDir}\textbf{C.} and \textbf{A.}).

\newpage

\newpage
\section{Results for the two-directional two-patch model Eq. \ref{model_equal}}

\subsection{Asymptotic analysis results for the two-directional two-patch model Eq. \ref{model_equal}}
We next investigated the long-term behavior of model Eq. \ref{model_equal}, where HBV is seeded in nodular structure 1 and moves at the same rate $\phi$ between the two patches. Model Eq. \ref{model_equal} has at most four \color{black} equilibria. The infection-free equilibrium:
\begin{align*}
E_4=(T_1^4,I_1^4,V_1^4,T_2^4,I_2^4,V_2^4) = (\frac{s_1}{d},0,0,\frac{s_2}{d},0,0),
\end{align*}
represents complete viral clearance in both patches, as before. The other three potential equilibria are of the form: $$E_j=(T_1^j,I_1^j,V_1^j,T_2^j,I_2^j,V_2^j),$$ $j\in\{5,6,7\}$, where
\begin{align*}
    I_1^j &=\frac{\phi T_1^j(-T_1^j\beta^2p^2s_2 + T_1^j\beta c d \delta p + T_1^j\beta d \delta p \phi + \beta c \delta p s_2+\beta \delta p \phi s_2 - c^2d\delta^2 - 2cd\delta^2\phi)}{\delta(\beta pT_1^j - c\delta - \delta\phi)(T_1^j\beta c p + \theta_ j\beta p \phi - c^2\delta - 2c\delta \phi)},\\
    V_1^j &=\frac{\phi(-T_1^j\beta^2p^2s_2+T_1^j \beta c d \delta p + T_1^j \beta d \delta p \phi + \beta c \delta p s_2 + \beta \delta p \phi s_2 - c^2d\delta^2 - 2cd\delta^2\phi)}{\beta(T_1^j\beta c p + T_1^j\beta p \phi - c^2 \delta - 2c\delta \phi)(\beta pT_1^j - c\delta - \delta \phi)},\\
    T_2^j &=
\frac{\delta (T_1^j \beta c p + T_1^j\beta p \phi - c^2\delta - 2c\delta\phi)}{\beta p (\beta p T_1^j - c\delta -\delta \phi)},\\
I_2^j &=\frac{T_1^j\beta^2p^2s_2 - T_1^j\beta c d \delta p - T_1^j\beta d \delta p \phi - \beta c \delta p s_2 - \beta \delta p \phi s_2 + c^2d\delta^2 + 2cd\delta^2\phi}{\delta\beta p(\beta p T_1^j - c\delta - \delta \phi)},\\
V_2^j &= \frac{T_1^j \beta^2 p^2 s_2 - T_1^j \beta c d \delta p - T_1^j\beta d \delta p \phi - \beta c \delta p s_2 - \beta \delta p \phi s_2 + c^2d\delta^2 + 2cd\delta^2\phi}{\delta \beta(T_1^j\beta c p + T_1^j\beta p \phi - c^2\delta - 2c\delta \phi)}.
\end{align*}
Here $T_1^j$ is the root of
\begin{equation}\label{poly}
    \gamma_0 \left(T_1^j\right)^3 + \gamma_1 \left(T_1^j\right)^2 + \gamma_2 \left(T_1^j\right)+\gamma_3=0,
\end{equation}
where
\begin{align*}
    \gamma_0 &= \beta dp^2(\beta c+\phi),\\
    \gamma_1 &= -\beta^2cp^2s_1-\beta^2p^2\phi s_1-\beta^2p^2\phi s_2 - 2\beta c^2d\delta p-3\beta c d \delta p \phi,\\
    \gamma_2 &= 2\beta c^2\delta p s_1 + 4\beta c \delta p \phi s_1 + \beta c \delta p \phi s_2 + \beta \delta p \phi^2 s_1 + \beta \delta p \phi^2 s_2 + c^3 d \delta^2 + 2c^2 d \delta^2 \phi,\\
    \gamma_3 &
    = -c\delta^2s_1(c^2+ 3c\phi+2\phi^2).
\end{align*}

Since $\gamma_0>0$, $\gamma_1<0$, $\gamma_2>0$ and $\gamma_3<0$, by Descartes' rule of signs, we can have one or three $T_1^j$ positive roots. Additionally, for the chronic equilibrium $E_j$ to exist, we need to guarantee positivity of the other variables in equilibrium $E_j$. This occurs when:

\begin{equation}
T_1^j<\min\big\{\frac{\delta(c+\phi)}{\beta p}, \frac{c\delta(c+2\phi)}{\beta p(c+\phi)}, \frac{\delta}{\beta p}\frac{\beta p s_2(c+\phi)-cd\delta(c+2\phi)}{\beta p s_2-d\delta(c+\phi)}\big\}.
\end{equation}

\begin{proposition} The infection-free equilibrium $E_4$ is locally asymptotically stable if
\begin{equation*}\label{R0_2D}
    \mathcal{R}_0^{2D} = \frac{\beta p s_2}{d\delta}\frac{p s_1\beta + d \delta (c+\phi)}{ps_1\beta(c+\phi) + d c\delta(c+2\phi)}<1,
\end{equation*}
and unstable otherwise.
\proof
We linearize Eq. \ref{model_equal} at the infection-free equilibrium $E_4$:
\begin{equation*}
\frac{d}{dt}
    \begin{bmatrix}
        T_1\\I_1\\V_1\\T_2\\I_2\\V_2
    \end{bmatrix}(E_4)=\begin{bmatrix}
        -d & 0 & -\frac{\beta s_1}{d} & 0 & 0 & 0\\
        0 & -\delta & \frac{\beta s_1}{d} & 0 & 0 & 0\\
        0 & p & -c-\phi & 0 & 0 & \phi\\0 & 0 & 0 & -d & 0 & -\frac{\beta s_2}{d}\\0 & 0 & 0 & 0 & -\delta & \frac{\beta s_2}{d}\\
        0 & 0 & \phi & 0 & p & -c-\phi
    \end{bmatrix}\begin{bmatrix}
        T_1\\I_1\\V_1\\T_2\\I_2\\V_2
    \end{bmatrix}(E_4),
\end{equation*}
The corresponding Jacobian matrix at $E_4$ is:
\begin{equation*}
    J(E_4) = \begin{bmatrix}
        -d & 0 & -\frac{\beta s_1}{d} & 0 & 0 & 0\\
        0 & -\delta & \frac{\beta s_1}{d} & 0 & 0 & 0\\
        0 & p & -c-\phi & 0 & 0 & \phi\\0 & 0 & 0 & -d & 0 & -\frac{\beta s_2}{d}\\0 & 0 & 0 & 0 & -\delta & \frac{\beta s_2}{d}\\
        0 & 0 & \phi & 0 & p & -c-\phi
    \end{bmatrix},
\end{equation*}
and the characteristic equation for model Eq. \ref{model_equal} at $E_4$ is:
\begin{align*}
    0=P_{4}(E_4)=\det(J(E_4)-\lambda \mathbbm{1}_6) = (\lambda + d)^2\begin{vmatrix}
        -\delta-\lambda & -\frac{\beta s_1}{d} & 0 & 0\\
        p & -c-\phi-\lambda & 0 & \phi\\
        0 & 0 & -\delta - \lambda & \frac{\beta s_2}{d}\\
        0 & \phi & p & -c-\phi-\lambda
    \end{vmatrix}.
\end{align*}
Two eigenvalues, $\lambda_{1,2}=-d$ are always negative. The remaining polynomial is:
\begin{equation*}
P_{41}(\lambda) = \lambda^4 + a_1\lambda^3 + a_2\lambda^2 + a_3\lambda + a_4,
\end{equation*}
where
\begin{align*}
    a_1 &= 2 c+2 \delta +2 \phi,\\
    a_2 &= (c+\delta)^2+2\phi(c+2\delta)+\frac{c^2\delta}{c+\phi}+\frac{\beta^2 p^2 s1 s2}{d^2\delta(c+\phi)}+\frac{\beta p s_1(c+\phi)+dc\delta (c+2\phi)}{d(c+\phi)}(1-\mathcal{R}_0^{2D}),\\
    a_3 &= 2(c+\delta)(c+\phi)\delta+2c\phi\delta+\frac{(c+\phi+\delta)\beta^2 p^2 s_1 s_2}{d^2 \delta (c+\phi)}+\frac{\beta ps_1 (c+\phi)+dc\delta(c+2\phi)}{d(c+\phi)}(c+\phi+\delta)(1-\mathcal{R}_0^{2D}),\\
    a_4&=\frac{\beta ps_1 \delta(c+\phi)+dc\delta^2(c+2\phi)}{d}(1-\mathcal{R}_0^{2D}).
\end{align*}
By the Routh-Hurwitz criteria, polynomial $P_{41}(\lambda)$ has negative roots or roots with negative real parts when $a_i>0$ and $a_1 a_2 a_3 > a_3^2+ a_1^2 a_4$. If $\mathcal{R}_0^{2D}<1$, then all $a_i>0$.
Moreover, using Maple, we can show that $a_1a_2a_3 > a_3^2 + a_1^2a_4$ when $\mathcal{R}_0^{2D} < 1$ (see \textbf{Appendix A}). Thus, by the Routh-Hurwitz criteria, if $\mathcal{R}_0^{2D} < 1$, the infection-free equilibrium $E_4$ is locally asymptotically stable. In contrast, if $\mathcal{R}_0^{2D} > 1$, the infection-free equilibrium $E_4$ is unstable. This concludes the proof.
\qed
\end{proposition}
Lastly, since explicit forms of chronic equilibria  $E_j$ ($j={5,6,7}$) are hard to determine, we examined their existence and asymptotic stability numerically.

\subsection{Structural identifiability results for model Eq. \ref{model_equal}}
As with the one-directional two-patch model, we investigated the structural identifiability of model Eq. \ref{model_equal} under two assumptions: (i) viral load is measured in both patches and the initial conditions of the model Eq. \ref{model_equal} are either known or unknown; (ii) susceptible liver cells are measured in both patches and the initial conditions of the model Eq. \ref{model_equal} are either known or unknown.

\begin{proposition}
    Given measurements for the $V_1(t)$  and $V_2(t)$ variables, system Eq. \ref{model_equal} is unidentifiable when initial conditions are unknown. In particular, parameters $\pi_1=\{ \beta, c, d, \delta, \}$ are globally structurally identifiable, and parameters $\pi_2=\{s_{1}, s_{2}, p\}$ are unidentifiable. If, additionally, initial conditions are known, then all parameters are globally structurally identifiable.

\proof
The input-output equations for model Eq. \ref{model_equal}, given data for $y_1=V_{1}(t)$ and $y_2=V_{2}(t)$, are:
\begin{align*}
& f(y_1, y_2, \pi_1,\pi_2)= \frac{d^3 y_1}{dt^3} y_1 - \frac{d^2 y_1}{dt^2} \frac{d y_1}{dt} + \frac{d^2 y_1}{dt^2} y_1^2 \beta + \frac{d^2 y_1}{dt^2} y_1 (c + d + \delta + \phi) - \left( \frac{d y_1}{dt} \right)^2 (c + \delta + \phi) \\&  + \frac{d y_1}{dt} \frac{d y_2}{dt} \phi + \frac{d y_1}{dt} y_1^2 \beta (c + \delta + \phi)
 + \frac{d y_1}{dt} y_1 d (c + \delta + \phi) + \frac{d y_1}{dt} y_2 \delta \phi - \frac{d^2 y_2}{dt^2} y_1 \phi
 - \frac{d y_2}{dt} y_1^2 \beta \phi \\
 &- \frac{d y_2}{dt} y_1 \phi (d + \delta) + y_1^3 \beta \delta (c + \phi)  - y_1^2 y_2 \beta \delta \phi + y_1^2 \left( - \beta p s_1 + c d \delta + d \delta \phi \right) - y_1 y_2 d \delta \phi=0,\\
& g(y_1, y_2, \pi_1, \pi_2)= - \frac{d^2 y_1}{dt^2} y_2 \phi + \frac{d y_1}{dt} \frac{d y_2}{dt} \phi - \frac{d y_1}{dt} y_2^2 \beta \phi - \frac{d y_1}{dt} y_2 \phi (d + \delta) + \frac{d^3 y_2}{dt^3} y_2 - \frac{d^2 y_2}{dt^2} \frac{d y_2}{dt}\\
& + \frac{d^2 y_2}{dt^2} y_2^2 \beta + \frac{d^2 y_2}{dt^2} y_2 (c + d + \delta + \phi)- \left( \frac{d y_2}{dt} \right)^2 (c + \delta + \phi) + \frac{d y_2}{dt} y_1 \delta \phi + \frac{d y_2}{dt} y_2^2 \beta (c + \delta + \phi) \\&+ \frac{d y_2}{dt} y_2 d (c + \delta + \phi) - y_1 y_2^2 \beta \delta \phi - y_1 y_2 d \delta \phi + y_2^3 \beta \delta (c + \phi) + y_2^2 \left( - \beta p s_2 + c d \delta + d \delta \phi \right)=0.
\end{align*}
Parameters $\{\pi_1,\pi_2\}$ are globally structurally identifiable if:
$$f(y_1, y_2, \pi_1, \pi_2)=f(y_1, y_2, \hat{\pi}_1, \hat{\pi}_2) \textrm{ and } g(y_1, y_2, \pi_1, \pi_2)=g(y_1, y_2, \hat{\pi}_1, \hat{\pi}_2) \textrm{ imply } \pi_1=\hat{\pi}_1 \textrm{ and } \pi_2=\hat{\pi}_2.$$
It is easy to see that:
$$\{\beta=\hat{\beta}\} \textrm{ , } \{d=\hat{d}\} \textrm{ , } \{\delta=\hat{\delta}\} \textrm{ , } \{c=\hat{c}\} \textrm{ , } \{\phi=\hat{\phi}\} \textrm{ , } \{p \times s_1=\hat{p}\times\hat{s}_1\} \textrm{ , } \{p \times s_2=\hat{p}\times\hat{s}_2\}.$$
Therefore $\{\beta, c, d, \delta, \phi\}$ are globally structurally identifiable and $\{s_1, s_2, p\}$ are unidentifiable. When initial conditions are known, we have
$$\frac{dV_{1}}{dt}(0)=p I_1(0) -cV_1(0)-\phi V_1(0)+\phi V_2(0).$$
As long as $I_1(0)\neq 0$, $p$ (and by default $s_1$ and $s_2$) are globally structurally identifiable. Hence, model Eq. \ref{model_equal} is globally structurally identifiable for known initial conditions. A summary of the results is given in Table \ref{tab:Identifiability_twoPatch}.
\qed
 \end{proposition}

\begin{proposition}
    Given measurements for the $T_1(t)$  and $T_2(t)$ variables, system Eq. \ref{model_equal} is unidentifiable when initial conditions are unknown. In particular, parameters $\pi_1=\{c, d, \delta, \phi, s_{1}, s_{2}\}$ are globally structurally identifiable, and parameters $\pi_2=\{\beta, p\}$ are unidentifiable. If, additionally, initial conditions are known, then all parameters are globally structurally identifiable.

\proof
The input-output equations for model Eq. \ref{model_equal}, given data for $y_1=T_{1}(t)$ and $y_2=T_{2}(t)$, are:
\begin{align*}
 & f(y_1,y_2, \pi_1, \pi_2)= - \frac{d^3 y_1}{dt^3} y_1^2 y_2^2 + 3 \frac{d^2 y_1}{dt^2} \frac{d y_1}{dt} y_1 y_2^2 - \frac{d^2 y_1}{dt^2} y_1^2 y_2^2 (c + \delta + \phi)
 - \frac{d^2 y_1}{dt^2} y_1 y_2^2 s_1  - 2 \left( \frac{d y_1}{dt} \right)^3 y_2^2 \\
&+ \frac{d y_1}{dt}^2 y_1 y_2^2 (c + \delta + \phi) + 2 \frac{d y_1}{dt}^2 y_2^2 s_1 + \frac{d y_1}{dt} y_1^3 y_2^2 \beta p - \frac{d y_1}{dt} y_1^2 y_2^2 \delta (c + \phi)
 - \frac{d y_1}{dt} y_1 y_2^2 s_1 (c + \delta + \phi) + \frac{d^2 y_2}{dt^2} y_1^3 y_2 \phi  \\ & - \left( \frac{d y_2}{dt} \right)^2 y_1^3 \phi + \frac{d y_2}{dt} y_1^3 y_2 \delta \phi + \frac{d y_2}{dt} y_1^3 \phi s_2  + y_1^4 y_2^2 \beta d p - y_1^3 y_2^2 \left( \beta p s_1 + c d \delta \right)  - y_1^3 y_2 \delta \phi s_2 + y_1^2 y_2^2 \delta s_1 (c + \phi)=0,\\
& g(y_1,y_2, \pi_1, \pi_2)= \frac{d^2 y_1}{dt^2} y_1 y_2^3 \phi - \left( \frac{d y_1}{dt} \right)^2 y_2^3 \phi + \frac{d y_1}{dt} y_1 y_2^3 \delta \phi + \frac{d y_1}{dt} y_2^3 \phi s_1 - \frac{d^3 y_2}{dt^3} y_1^2 y_2^2 + 3 \frac{d^2 y_2}{dt^2} \frac{d y_2}{dt} y_1^2 y_2 \\
& - \frac{d^2 y_2}{dt^2} y_1^2 y_2^2 (c + \delta + \phi) - \frac{d^2 y_2}{dt^2} y_1^2 y_2 s_2 - 2 \left( \frac{d y_2}{dt} \right)^3 y_1^2 + \left( \frac{d y_2}{dt} \right)^2 y_1^2 y_2 (c + \delta + \phi) + 2 \left( \frac{d y_2}{dt} \right)^2 y_1^2 s_2 + \frac{d y_2}{dt} y_1^2 y_2^3 \beta p \\
&- \frac{d y_2}{dt} y_1^2 y_2^2 \delta (c + \phi) - \frac{d y_2}{dt} y_1^2 y_2 s_2 (c + \delta + \phi) + y_1^2 y_2^4 \beta d p - y_1^2 y_2^3 (\beta p s_2 + c d \delta) + y_1^2 y_2^2 \delta s_2 (c + \phi) - y_1 y_2^3 \delta \phi s_1=0.
\end{align*}
Parameters $\{\pi_1,\pi_2\}$ are globally structurally identifiable if:
$$f(y_1, y_2, \pi_1, \pi_2)=f(y_1, y_2, \hat{\pi}_1, \hat{\pi}_2) \textrm{ and } g(y_1, y_2, \pi_1, \pi_2)=g(y_1, y_2, \hat{\pi}_1, \hat{\pi}_2) \textrm{ imply } \pi_1=\hat{\pi}_1 \textrm{ and } \pi_2=\hat{\pi}_2.$$
It is easy to see that:
$$\{s_1=\hat{s}_1\} \textrm{ , } \{s_2=\hat{s}_2\} \textrm{ , } \{\delta=\hat{\delta}\} \textrm{ , } \{d=\hat{d}\} \textrm{ , } \{c=\hat{c}\} \textrm{ , } \{\phi=\hat{\phi}\} \textrm{ , } \{\beta\times p=\hat{\beta}\times \hat{p}\}.$$
Therefore $\{s_1, s_2, d, \delta, c, \phi\}$ are globally structurally identifiable and $\{\beta, p\}$ are unidentifiable. When initial conditions are known, we have:
$$\frac{dT_{1}}{dt}(0)=s_1-d T_1(0)-\beta T_1(0) V_1(0),$$
making $\beta$ (and by default $p$) globally structurally identifiable. Hence, model Eq. \ref{model_equal} is globally structurally identifiable for known initial conditions. A summary of the results is given in Table \ref{tab:Identifiability_twoPatch}.
\qed
 \end{proposition}

 \begin{table}[h!]
\centering
\begin{tabular}{|p{1.7in} | p{1.7in}| p{1.8in}|}
  \hline
  Observe States & Initial Conditions Known & Initial Conditions Unknown\\ \hline
Model with $V_{1}(t)$ and $V_{2}(t)$ data& Globally structurally identifiable \{$\beta$, c, d, $\delta$, $\phi$, p, $s_{1}$, $s_{2}$\} & Globally structurally identifiable \{$\beta$, c, d, $\delta$, $\phi$\}; Unidentifiable \{$s_{1}$, $s_{2}$, p\}; Correlations $\{s_1\times p=\hat{s}_1 \times \hat{p}\}$, $\{s_2\times p=\hat{s}_2 \times \hat{p}\}$ \\\hline
Model with $T_{1}(t)$ and $T_{2}(t)$ data& Globally structurally identifiable \{$\beta, c, d, \delta, \phi, p, s_{1}, s_{2}$\} & Globally structurally identifiable \{$c, d, \delta, \phi, s_{1}, s_{2}$\}; Unidentifiable \{$\beta, p$\}; Correlations $\{\beta\times p=\bar{\beta}\times \bar{p} \}$   \\ \hline
\end{tabular}
\caption{Identifiability analysis for model Eq. \ref{model_equal}, performed using the DAISY software \cite{bellu2007daisy}. For the simulations with initial conditions, all initial conditions are known.
}
\label{tab:Identifiability_twoPatch}\end{table}

\subsection{Numerical results for model Eq. \ref{model_equal}}
As seen above, under known initial conditions and unlimited noise-free measurements of both $V_1(t)$ and $V_2(t)$ (or $T_1(t)$ and $T_2(t)$), all parameters of model Eq. \ref{model_equal} can be identified from data.

As before, we assumed that the entire hepatocyte population $s=d\times (T_1(0)+T_2(0))$ is susceptible to HBV infection and considered three cases for the susceptible cells ratio within the two patches $s_1:s_2 =\{10:90, 50:50, 90:10\}$ (see Table \ref{tab:Param_twoPatch_twoDirection}). As before, we set $\delta=d=0.01$ /day \cite{ciupe2007modeling}, fixed $c=4.4$ /day and fitted parameters $\{\beta, p, \phi\}$ using the \texttt{fminsearchbnd} function in Matlab2021a (see \textbf{Materials and Methods} section for details). A summary of estimates for each case is given in Table \ref{tab:Param_twoPatch_twoDirection} and the model dynamics are plotted in Fig. \ref{fig:twoPatch_twoDirection}.

As with the one-directional two-patch model, we obtain similar infectivity rates among the three patch recruitment cases, \textit{i.e.} $\beta=2.93\times 10^{-9}$ ml/(virion $\times$ day), $\beta=2.96\times 10^{-9}$ ml/(virion $\times$ day), and $\beta=2.94\times 10^{-9}$ ml/(virion $\times$ day), for \textbf{cases 1, 2} and \textbf{3}, respectively. Moreover, viral production rates are similar among the three patch recruitment cases, \textit{i.e.} $p=1053$ virion/(ml$\times$ day), $p=1055$ virion/(ml$\times$ day), and $p=1049$ virion/(ml$\times$ day), respectively. The estimate of the movement rate is 50-times higher in \textbf{case 1} and \textbf{case 3}, compared to \textbf{case 2}, \textit{i.e.} $\phi=5$ /day versus $\phi=0.1$ /day. To determine if the results can be generalized when we add noise to the data, we performed practical identifiability analysis using Monte Carlo approaches (see \textbf{Materials and Methods} for details). We found that all parameters are strongly practically identifiable for all cases, with the exception of $\phi$, which is weakly practically identifiable in all cases (see Table \ref{tab:Practical_TwoDir}).

\begin{table}[h!]
\begin{tabular}{|c||c|c|c||c|c|c||c|c|c||}
\hline
\multicolumn{1}{|c||}{} & \multicolumn{3}{|c||}{\textbf{case 1}} & \multicolumn{3}{|c||}{\textbf{case 2}} & \multicolumn{3}{|c||}{\textbf{case 3}} \\
\hline ARE & $\beta$ &$\phi$  & $p$ & $\beta$ &$\phi$  & $p$ & $\beta$ &$\phi$  & $p$ \\
\hline 0\% & 0 & 0 & 0 & 0 & 0 & 0 & 0 & 0 & 0 \\
\hline 1\% & 0.0207 & 1.8487 & 0.3372 & 0.0242 & 10.3416 & 0.3967 & 0.0366 & 2.5045 & 0.5624 \\
\hline 5\% & 0.2271 & 3.8460 & 3.2488 & 0.1183 & 76.3263 & 1.9011 & 0.3342 & 1.4423 & 4.6259 \\
\hline 10\% & 0.4226 & 2.3097 & 5.9290 & 0.2775 & 64.4378 & 4.2071 & 0.4965 & 1.5473 & 6.8888 \\
\hline 20\% & 0.6925 & 1.4294 & 9.7573 & 0.6096 & 51.1635 & 8.8355 & 0.7205 & 1.6928 & 10.0169 \\
\hline 30\% & 0.8235 & 1.4597 & 11.9580 & 0.8160 & 36.3667 & 11.7393 & 0.8739 & 1.5403 & 12.5380 \\ \hline
Identifiable? & Yes & Weakly & Yes & Yes & Weakly & Yes & Yes & Weakly & Yes    \\
\hline
\end{tabular}
\caption{MC approach for the two-directional two-patch model Eq. \ref{model_equal}. }\label{tab:Practical_TwoDir}
\end{table}

In all three cases we are in the $R_0^{2D}>1$ regime ($R_0^{2D}=22$, for \textbf{case 1}; $R_0^{2D}=23.6$, for \textbf{case 2}; and $R_0^{2D}=2.3$, for \textbf{case 3}). While we know from the analytical results that this corresponds to instability of the clearance equilibrium $E_4$, we do not know if it guarantees the stability of the both-patches viral persistence equilibrium $E_5$ (note that equilibrium solutions for single-patch viral persistence do not exist for this model). We investigated these long-term results numerically by deriving one- and two-dimensional bifurcation diagrams.

\begin{figure}[h!]
    \centering
    \includegraphics[width=\linewidth]{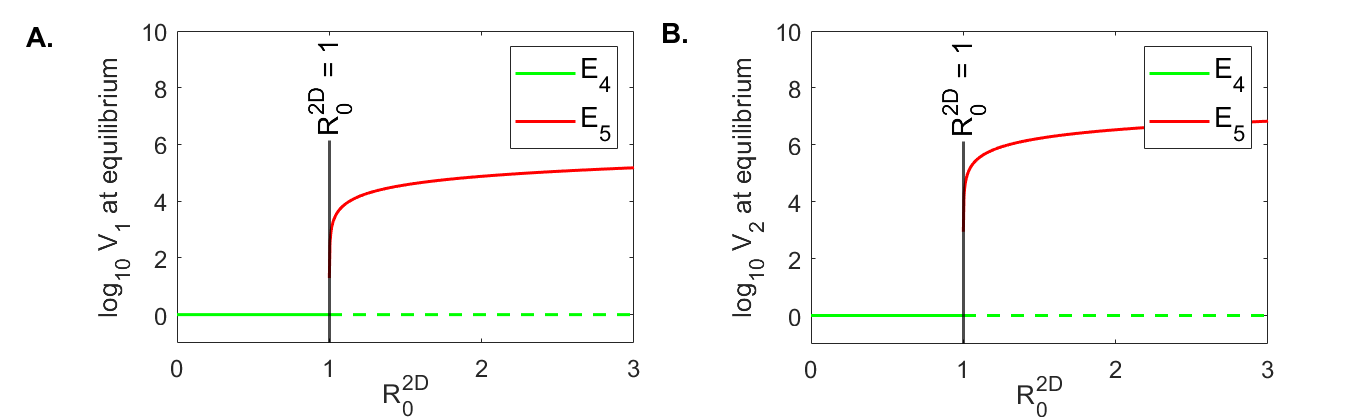}
    \caption{One dimensional bifurcation diagram for model Eq. \ref{model_equal}: $\log_{10} V_1$ and $\log_{10} V_2$ at equilibrium versus $R_0^{2D}$. The other parameters are given in Table \ref{tab:Param_twoPatch_oneDirection} for \textbf{case 2}.}
    \label{fig:1DBif_twoDir}
\end{figure}

\begin{figure}[h!]
    \centering
       \includegraphics[width=0.32\linewidth]{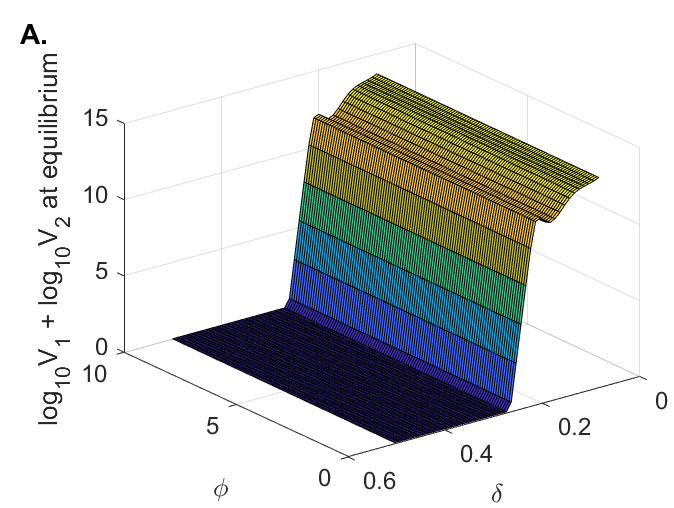} \includegraphics[width=0.32\linewidth]{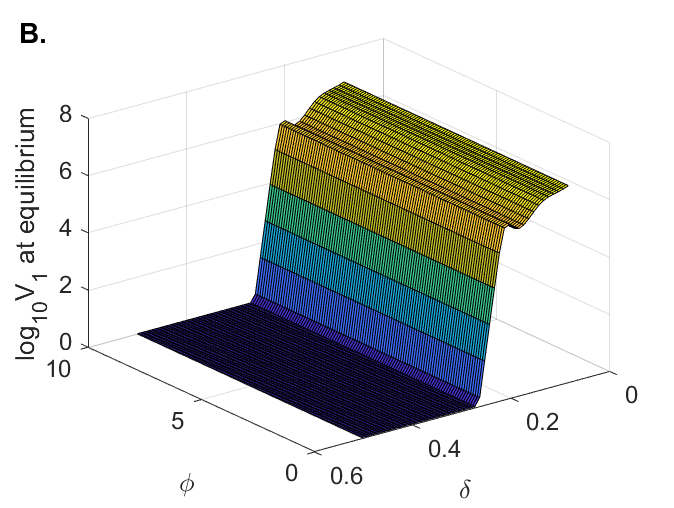}
    \includegraphics[width=0.32\linewidth]{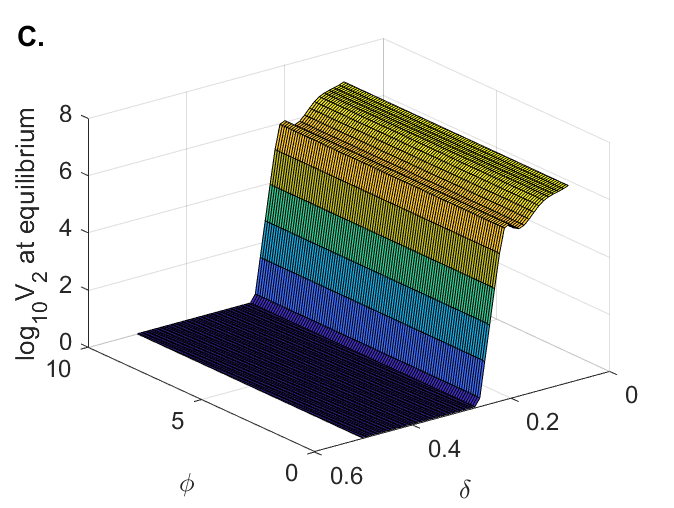}

    \caption{Two dimensional bifurcation diagram for model Eq. \ref{model_equal}: \textbf{A.} $\log_{10} V_1+\log_{10} V_2$ at equilibrium; \textbf{B.} $\log_{10} V_1$ at equilibrium; \textbf{C.} $\log_{10} V_2$ at equilibrium versus $\delta$ and $\phi$. The other parameters are as in Table \ref{tab:Param_twoPatch_twoDirection}, case 2.}
    \label{fig:2D_TwoDir}
\end{figure}

We created one-dimensional bifurcation diagrams as follows. We numerically solved system Eq. \ref{model_equal} for parameters in \textbf{case 2} (Table \ref{tab:Param_twoPatch_twoDirection}) and $\delta\in(0.01, 0.6)$ using the \texttt{ode15} function in Matlab2021a. We ran the model for $t_{end}=3000$ days and plotted $V_1(t_{end})$ and $V_2(t_{end})$ versus $R_0^{2D}$. We found that both $V_1$ and $V_2$ are cleared when $R_0^{2D}<1$ and persist when $R_0^{2D}>1$ (Fig. \ref{fig:1DBif_twoDir}\textbf{A.} and \textbf{B.}). We also created two-dimensional bifurcation diagrams (using the numerical values of $V_1(t_{end})$ and $V_2(t_{end})$, as before), for varied values of $\delta\in (0.01, 0.6)$, and $\phi\in(0,10)$. We found little variability in the long-term dynamics of $V_1$ and $V_2$, with both viruses being cleared for high infected cells killing rate $\delta$ and persisting for low infected cells killing rate $\delta$, regardless of the movement parameter $\phi$ (Fig. \ref{fig:2D_TwoDir}). This implies that, when virus moved between patches, immune mediated removal of infected cells is mandatory for viral clearance.

\section{Discussion and conclusion}
In this study, we developed two within-host mathematical models of HBV infection that take into account the abnormal nodular structure of the liver following prolonged hepatitis B infection and disease \cite{yuan2019hbv, bataller2005liver, roehlen2020liver}. For simplicity, we only modeled a two-patch liver structure and assumed that: (1) each patch has a different numbers of cells susceptible to HBV infection, (2) HBV is seeded in a single patch, (3) all other virus-host interactions are patch-agnostic. The first model, given by Eq. \ref{model_oneD}, assumes one-directional hepatitis B virus movement between patches. The second model, given by Eq. \ref{model_equal}, assumes two-directional hepatitis B virus movement between patches. We used the models to determine differences in overall hepatitis B virus infection under viral localization (as given by the one-directional model Eq. \ref{model_oneD}) and free viral movement between patches (as given by the two-directional model Eq. \ref{model_equal}).

We analyzed both models using asymptotic stability techniques and found that when HBV movement is one-directional, as given by model Eq. \ref{model_oneD}, equilibrium dynamics include viral clearance from both patches, viral clearance from the HBV seeding patch alone, and virus persistence in both patches. When HBV movement is two-directional, as given by model Eq. \ref{model_equal}, however, equilibrium dynamics only include viral clearance from both patches, and virus persistence in both patches. In other words, competitive exclusion and viral containment within parts of the liver only happen when HBV movement between patches is irreversible. This can be a result of collagen production, excessive
accumulation of extracellular matrix, and liver dysfunction \cite{bataller2005liver}.

We fitted the models to HBV DNA data from one immunocompromised, humanized mice  \cite{zhang2023replication} and estimated three key parameter values for each model: the infectivity rate $\beta$, the viral production rate $p$, and the HBV movement rate between patches $\phi$. Since empirical data (from an immunocompromised animal) showed viral persistence following the initial expansion, we ignored the effect of immune-mediated killing rate of infected cells $\delta$ (which was set to uninfected cell death rate) and relaxed this assumption later on. We considered differences in susceptibility to HBV infection between patches, by varying the target cell recruitment ratio, $s_1:s_2$ between 10:90, 50:50 and 90:10, while keeping the overall liver cell numbers fixed (a reasonable assumption in an immunocompromised mice \cite{wieland2005stealth}).

We found that the estimates for the infectivity rate $\beta$ and for the viral production rate $p$ are not affected (in a significant manner) by the target cell recruitment ratio $s_1:s_2$ or by the ability of HBV to move back-and-forth between patches ($\phi_{21}=0$ or $\phi_{21}=\phi_{12}=\phi$). The movement rate $\phi$, however, is significantly impacted by the
target cell recruitment ratio $s_1:s_2$ and by the model. In particular, for the one-directional two-patch movement model Eq. \ref{model_oneD}, HBV chooses movement towards the non-founding patch at significantly higher rates (up to 50-times higher) when the initial patch has more susceptible cells. This outcome may be due to larger numbers of HBV being created in the original patch in the 50:50 and 90:10 cases, resulting in larger cell movement. By contrast, for the two-directional two-patch movement model Eq. \ref{model_equal}, HBV chooses higher movement rates (up to 50-times higher) when the patches have unbalanced numbers of susceptible cells (10:90 and 90:10, respectively) and limited movement for the balanced patches (50:50). This outcome may be due to each patch having a sufficient number of target cells available for HBV infection, limiting viral diffusion.

For the one-directional model Eq. \ref{model_oneD}, the equilibrium virus levels in patch 2 exceed equilibrium virus levels in patch 1, regardless of the target cell susceptibility ratio (Fig. \ref{fig:twoPatch_oneDirection}). By contrast, for the two-directional model Eq. \ref{model_equal}, equilibrium virus levels in patch 2 exceed those in patch 1 for the $10:90$ target cell susceptibility, but are equal to and lower than those in patch 1 for the $50:50$ and $90:10$ target cell susceptibility ratio (Fig. \ref{fig:twoPatch_twoDirection}). Therefore, the quantitative outcomes are influenced by the ability of the virus to move one- or two-directionally between patches. Our results are not influenced by identifiability issues, with both models being structurally (globally, for known initial conditions) and practically (at least weakly) identifiable.

To determine how the results change in an immune competent host, we derived one- and two-dimensional bifurcation diagrams by using $\delta$ (the killing rate of infected cells) as bifurcation parameter. We varied $\delta$ from $0.01$ /day (corresponding to infected cells lifespan of 100 days) to $0.6$ /day (corresponding to infected cells lifespan of 1.7 days).

For model Eq. \ref{model_oneD} we found, as expected, that increasing $\delta$ results in $R_0^{1D}$ decrease. When $R_0^{1D}$ decreases below 1, we observe viral clearance from both patches (Fig. \ref{fig:1DBif_oneDir}, solid green lines). Interestingly, viral clearance from both patches requires that $\delta>0.45$ /day (lifespan of infected cells of 2.2 days or lower) when $s_1:s_2=10:90$, $\delta>0.24$ /day (lifespan of infected cells of 4 days or lower) when $s_1:s_2=50:50$, and $\delta>0.055$ /day (lifespan of infected cells of 18 days or lower) when $s_1:s_2=90:10$. This suggests that higher immune responses are necessary for viral clearance when the majority of susceptible cells are in patch 2. By contrast, competitive exclusion outcomes (resulting in viral clearance in patch 1, and persistence in patch 2) occur for $\delta \in (0.05, 0.45)$ when $s_1:s_2=10:90$, $\delta\in(0.12, 0.24)$ when $s_1:s_2=50:50$, and not at all (in the ranges considered) when $s_1:s_2=90:10$
(Fig. \ref{fig:1DBif_oneDir}, solid red lines and Fig. \ref{fig:2D_oneDir}). This suggests a limited range of immune responses lead to competitive exclusion when the majority of susceptible cells are in patch 1.

For model Eq. \ref{model_equal} we found that viral clearance from both patches requires that $\delta>0.27$ /day (lifespan of infected cells of 3.7 days or lower) when $s_1:s_2=10:90$; $\delta>0.234$ /day (lifespan of infected cells of 4.27 days or lower) when $s_1:s_2=50:50$; and $\delta>0.023$ /day (lifespan of infected cells of 43 days or lower) when  $s_1:s_2=90:10$. This suggests that higher immune responses are necessary for viral clearance when the susceptible cells in patch 1 equal or exceed the susceptible cells in patch 2 (Fig. \ref{fig:2D_TwoDir}).

Our study has several limitations. First, the models ignore liver proliferation following liver stress and death  \cite{summers2003hepatocyte}. We accounted for liver proliferation by choosing different numbers of susceptible cells in each patch. Additional data is needed to allow for increase in model complexity through the addition of a proliferation term. Second, we ignored cure of infected cells by non-cytolytic processes \cite{guidotti1999viral, wieland2004expansion}, and intracellular viral effects \cite{ciupe2024incorporating} and assumed that the most important host feature is the immune-mediated killing of infected cells, $\delta$. It is known that HBV-induced liver disease is immune-mediated \cite{ciupe2007role, ciupe2014antibody, ciupe2024incorporating}, so more detailed models that account for immune populations and immune functions will be part of future work.
Third, we used an oversimplified model for the abnormal nodular liver structure following disease, that is composed of just two patches. While this allowed for analytical tractability of our results, a multi-patch model may result in richer dynamics and even patterns of infection. Lastly, to identify parameters, we had to assume that the initial conditions are known and $I_1(0)\neq 0$. While most within-host models assume that $I_1(0)=0$, we bypassed the initial infection and started with one infected cell $I_1(0)=1$, making parameter estimates dependent (in a non-significant way - results not shown) on this assumption.

In conclusion, we developed mathematical models that considered the formation of liver nodular structures following hepatitis B viral infection and disease and used them to investigate virus dynamics within-patches and systemically. We predicted different outcomes when viral movement between patches is either irreversible or reversible. Moreover, we found that cell susceptibility to infection within nodular structures, the movement rate between patches, and the immune-mediated infected cell killing have the most influence on the results.

\newpage

\begin{appendices}

\section{Stability of equilibrium $E_4$}
To finalize the proof of \textbf{Proposition 4.}, we need to show that $a_1 a_2 a_2>a_3^2+a_1^2 a_4$. We are including here the maple file showing that result.
\begin{figure}[h!]
    \centering
\includegraphics[width=.9\linewidth]{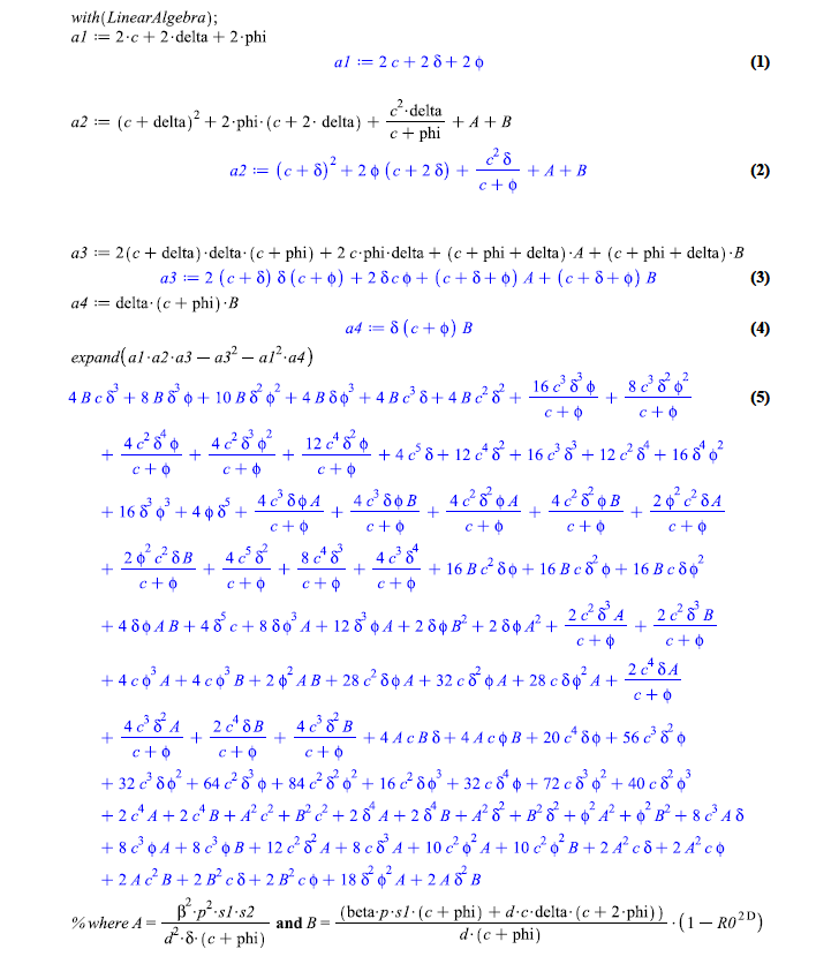}
 \caption{Supporting analysis for the stability of equilibrium $E_4$.}
\label{fig:Maple_E4}
\end{figure}

\end{appendices}

\newpage

\clearpage

\section*{Declarations}

\begin{itemize}
\item Funding. SMC acknowledges partial support from the NIH NIGMS grant 1R01GM152743-01 and National Science Foundation grant
No. 2051820. This research was enabled in part through the Virginia Tech Center for the Mathematics of Biosystems (VTCMB-033500).
\item Conflict of interest/Competing interests. We declare no conflict of interest associated with this publication.
\item Data availability. All data necessary to replicate the results in this article are available at \\ https://github.com/StancaCiupe/HBV-multi-patch.
\item Code availability. All code necessary to replicate the results in this article are available at \\ https://github.com/StancaCiupe/HBV-multi-patch.
\item Author contribution. Conceptualization: SMC; Methodology: KC, OS, SMC; Formal analysis and investigation: KC, OS, SMC; Writing- original draft preparation: SMC; Writing- review and editing: KC, OS, SMC; Funding acquisition: OS, SMC; Resources: OS, SMC; Supervision: OS, SMC.

\end{itemize}

\newpage
\bibliographystyle{unsrt}
\bibliography{References}

\end{document}